\documentclass[aps,twocolumn,nofootinbib,
preprintnumbers,superscriptaddress]{revtex4}

\usepackage{amsmath,amssymb,amsfonts,amsthm,graphicx, epsf,dsfont,dcolumn,yfonts}
\usepackage{latexsym}
\usepackage[colorlinks=true, pdfstartview=FitV, linkcolor=black, citecolor=black, urlcolor=black]{hyperref}

\usepackage{color}
\usepackage{slashed} 
\newcommand{\be}{\begin{equation}}
\newcommand{\ee}{\end{equation}}
\newcommand{\bea}{\begin{eqnarray}}
\newcommand{\eea}{\end{eqnarray}}
\def\ie{{\it i.e.~}}

\newcommand{\bwt}{\begin{widetext}}
\newcommand{\ewt}{\end{widetext}}

\def\det{{\rm{det}}}

\def\G{\Gamma}

\def\ads{{\rm AdS}}

\begin{document}

%\title{Cuprate Physics from Holographic Mott Insulators}
\title{Dynamical Gap and Cuprate-like Physics from Holography}

\author{Mohammad Edalati, Robert G. Leigh, Ka Wai Lo and  Philip W. Phillips}
%\vspace{25pt}
\affiliation{ Department of Physics,
University of Illinois at Urbana-Champaign, Urbana IL 61801, USA}
%\vspace{15pt}
%{\tt edalati@illinois.edu,}~~

\begin{abstract}
We study the properties of fermion correlators in a boundary
theory dual to the Reissner-Nordstr\"om AdS$_{d+1}$ background in the
presence of a bulk dipole (Pauli) interaction term with
strength $p$.  We show that by simply changing the value of the
parameter $p$ we can tune continuously from a Fermi liquid (small
$p$), to a
marginal Fermi liquid behavior at a critical value of $p$, to a generic
non-Fermi liquid at intermediate values of $p$, and finally to a Mott
insulator at large values of the bulk Pauli coupling.  As all of
these phases are seen in the cuprate phase diagram, the
 holographic model we study has the key elements of the strong
coupling physics typified by Mott systems.  In addition, we extend our
analysis to finite temperature and show that the Mott gap closes.  Of
particular interest is that it closes when the ratio of the gap to the
critical temperature is of the order of ten.  This behavior is very much similar to that observed in the classic Mott insulator VO$_2$.   We then analyze the non-analyticities of the boundary theory fermion correlators for generic values of frequency and momentum by calculating the quasi-normal modes of the bulk fermions. Not surprisingly, we find no evidence for the dipole interaction inducing an instability in the boundary theory. 
Finally, we briefly consider the introduction of superconducting condensates, and find that in that case, the fermion gap is driven by scalar-fermion couplings rather than by the Pauli coupling.

\end{abstract}

\maketitle

\section{Introduction}

Holography can offer unprecedented insight into the dynamics of
strongly coupled systems. In the recent past, it has become clear that
the domain of applicability of holography goes beyond high energy
physics and includes strongly-correlated sytems in condensed matter
physics, as well  (see \cite{Hartnoll:2009sz, Herzog:2009xv,
  McGreevy:2009xe,Horowitz:2010gk} for reviews). Since in most cases
of interest, we do not possess a microscopic understanding of the
field theory dynamics under study, a phenomenological point of view
(the so-called ``bottom-up" approach) is taken where a minimal
gravitational setup is devised for analyzing a specific
strong-coupling feature of a system. In some situations where string
or M-theory completion of a bottom-up construction is known,  one may
wonder how a bottom-up result  is modified in a
top-down approach. For example, a quantity of interest in
strongly-correlated condensed matter systems, which can easily be
computed using holography, is the fermion spectral function (which is
proportional to the imaginary part of the fermion retarded two-point
function). There has recently been much discussion about this quantity
in the holographic literature \cite{Lee:2008xf, Liu:2009dm,
  Cubrovic:2009ye, Faulkner:2009wj, Faulkner:2010tq} where analyzing
the Dirac equation for a charged probe fermion propagating in a
gravitational background (usually a charged black hole), one can show
that the retarded two-point function of the dual fermionic operator in
the boundary theory shows a variety of unexpected emergent phenomena. In top-down approaches to holographic systems, the fermions are generically coupled to gravity and gauge fields in a variety of ways, beyond minimal  coupling. It is certainly desirable to analyze how such non-minimal bulk  couplings modify the  fermion spectral function in the boundary theory, or may lead to new interesting emergent phenomena.

Recently, it was shown \cite{Gauntlett:2009zw, Cassani:2010uw, Gauntlett:2010vu,Liu:2010sa} that there are consistent truncations of ten and eleven dimensional supergravities to five and four dimensional bulk theories that possess an interesting class of gauge interactions and charged matter, allowing for novel condensed matter physics phenomena, such as superconductivity, to be explored in a consistent Ôtop-downÕ approach  \cite{Denef:2009tp, Gubser:2009qm, Gauntlett:2009dn}. The fermionic sector of these truncations has also been worked out in \cite{Bah:2010yt,Bah:2010cu} where a number of generic  couplings for the fermions (with possible applications to strongly-correlated condensed matter systems) have been realized. 

Motivated by these studies,  we considered in \cite{Edalati:2010ww}  a
generic non-minimal fermion coupling in which a spin-$1/2$ fermion
couples to the gauge field  through a dipole (Pauli) interaction of
the form $F_{ab}\bar\psi \Gamma^{ab}\psi$. In fact, we just considered
the simplest possible setup, in which a spin-$1/2$ fermion propagates
in the background of a Reissner-Nordstr\"om AdS$_{d+1}$  black
hole. We found that as one changes the strength of this interaction,
spectral weight of the dual fermionic operator is transferred between bands, and beyond a critical
value of the dipole coupling, a  gap emerges in the fermion density of
states. We then concluded that a possible interpretation of this
interaction is that it drives the dynamical formation of a (Mott) gap,
in the absence of continuous symmetry breaking.  In the Hubbard model, the Mott gap forms in
$d>1$ once the on-site interaction $U$ exceeds a critical value in the half-filled system.
Upon doping, spectral weight shifts from high to low energies.
Consequently, we argued that the strength of the dipole
interaction  mimics the combined effects of doping and the on-site
interaction strength $U$. 

\begin{figure}
\centering
\includegraphics[width=6.5cm]{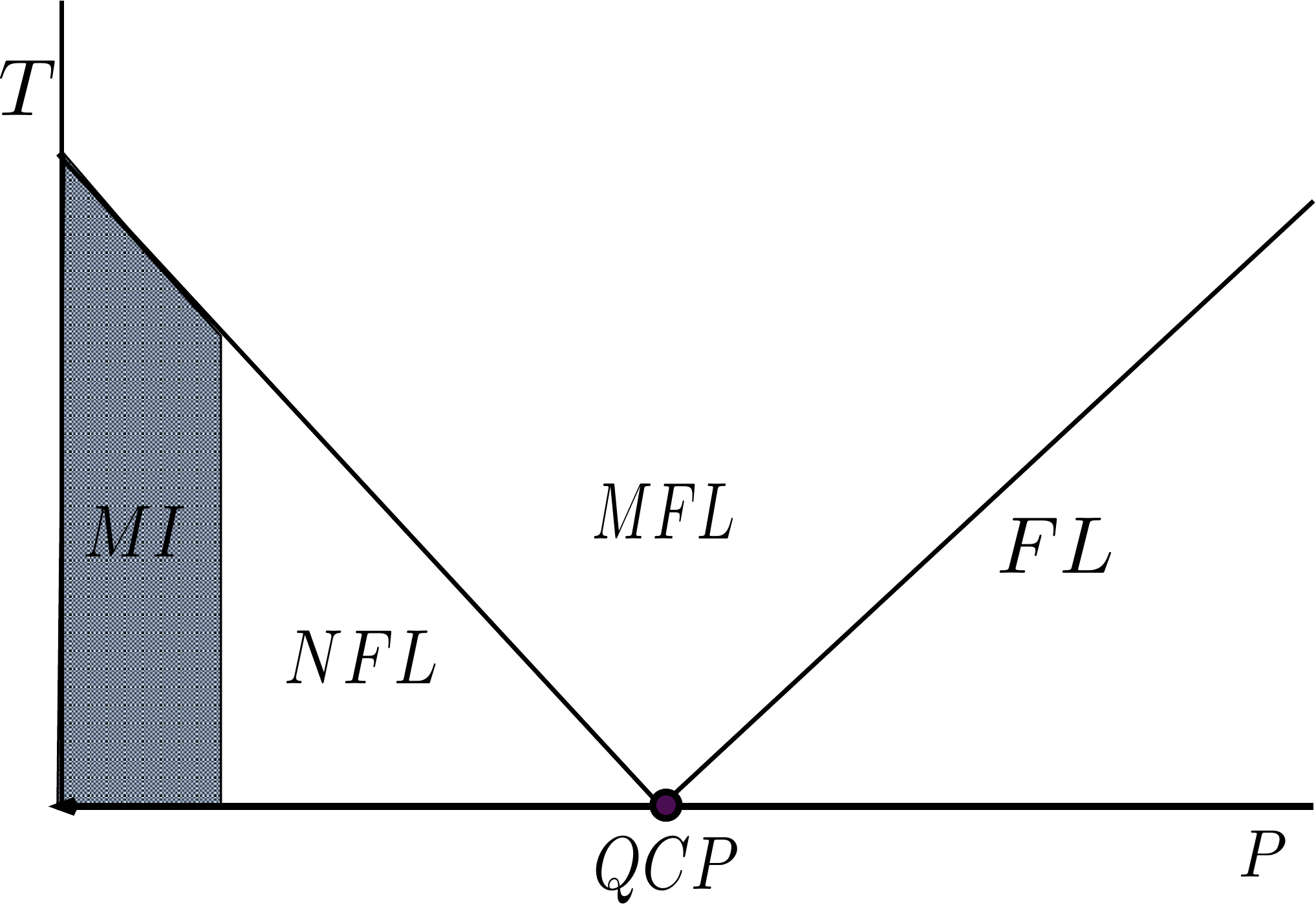}
\caption{\label{pdiag} \footnotesize{A cartoon of the phase diagram of the boundary theory considered  here.  MI indicates a Mott insulator, a phase with a gap in the absence of symmetry
  breaking.  NFL denotes non-fermi liquid behavior which is distinct from the gapped
  spectrum of a Mott insulator.  MFL indicates marginal-Fermi liquid behavior
  in which the ``electron" self energy scales as $\omega\log\omega$ at
  $T=0$, and a FL (Fermi liquid) regime in which the dispersion is
  linear in frequency.  The tuning parameter in this model is the Pauli (dipole) coupling. Similar behavior is obtained in the non-superconducting features of the cuprate materials by tuning the hole-doping level
  $x$.}}
\end{figure}

In this paper, we continue our study of the dipole interaction in more detail. 
We investigate the existence of Fermi surfaces as the
dipole coupling $p$ is varied. For the range of parameters considered in
this paper, we find that there is no Fermi surface above a certain
value of $p$. This is the range of parameters
considered in \cite{Edalati:2010ww}, where the boundary theory
exhibits two main features of Mott insulators, a dynamically generated
gap (in the absence of continuous symmetry breaking) and spectral weight transfer. In addition, for the values of the dipole coupling $p$ for which there is a Fermi surface in the boundary theory, we find that at a critical value of this coupling, marginal-Fermi liquid behavior ensues.  Consequently, we are able with our model (see
Figure \ref{pdiag}) to describe at one
extreme, the Mott insulating state with a dynamically generated gap, a
transition to a marginal Fermi liquid (MFL) and at the other, a Fermi-liquid-like regime
in which the excitation spectrum scales linearly with the frequency.
As all of these regimes are accessed simply by changing the value of
the dipole coupling $p$, this suggests a direct parallel between $p$ and
the hole-doping level in the high-temperature copper-oxide superconductors
(hereafter cuprates).  In the cuprates, the strong electron
correlation physics ends at a value of doping (typically optimal
doping) where
marginal-Fermi liquid \cite{varma} behavior ensues.  Perhaps the occurence of
the MFL state of the probe fermions in our holographic setup is an indication that
this phenomenological model is ultimately a robust feature of the transition from
strong to weakly interacting physics in doped Mott systems. 
 
Having discussed the pole structure of the  boundary theory fermion (retarded) correlators for non-zero values of $p$ at small frequency, we analyze the non-analyticities of those
correlators for generic values of frequency and momentum. We do this
by numerically calculating the quasi-normal modes of the
bulk fermion in the (extremal) Reissner-Nordstr\"om AdS$_{d+1}$
background. Following the motion of the poles in the complex frequency plane as a function of momentum, we compute their dispersion
relations and, for the range of parameters considered, confirm that all of the poles stay in the lower half of
the complex frequency plane, for all momenta. Hence, as expected,
turning on a non-zero bulk dipole coupling in our set up does not
cause an instability in the boundary theory. 
%We also study the
%effects of the bulk dipole interaction on fermion correlators
%in a boundary theory which is dual to a charged black hole background
%whose entropy vanishes at zero temperature, and find that  the bulk
%dipole interaction has qualitatively the same effects on fermion
%correlators as it did in the case of the boundary theory dual to the
%Reissner-Nordstr\"om AdS$_{d+1}$ background. 
%This indicates that our results seem to be robust to a holographic  setup with zero entropy at zero temperature.

 We consider our setup at finite temperature and find that as the
temperature increases the gap closes and, moreover, the critical
temperature for which this happens is much less than the value of the
gap.  In this regard, the closing of the gap induced by temperature
parallels what one obtains in the classic Mott system VO$_2$
\cite{zylbert}, namely, the gap closes at a temperature much lower
than the gap.

The paper is organized as follows. In section \ref{sectiontwo} we consider a bulk fermion in the Reissner-Nordstr\"{om} AdS$_{d+1}$ black hole background and couple it to the U(1) gauge field non-minimally through a dipole interaction with strength $p$. We then derive the Dirac equations and 
%show non-zero $p$ modifies the scaling of the fermion operators in the infrared by shifting the momentum. We 
rewrite them as flow equations which will be more convenient for numerically calculating the boundary theory fermion correlators. In section \ref{sectionthree} we investigate the existence of Fermi surfaces as a function of $p$ by solving the Dirac equations at zero frequency.  
%For the values of the mass and the charge (of the bulk fermion)  considered there, we find that  beyond a critical value of $p$ there is no Fermi surface. 
In section \ref{sectionfour}, we first discuss the small-frequency behavior of the poles of the boundary theory fermion (retarded) correlators when $p$ is non-zero.  We then analyze the non-analyticities of those correlators for generic values of frequency and momenta. In section \ref{sectionfive} we study the effects of temperature in our holographic setup.
% Section \ref{sectionsix} discusses how the bulk dipole interaction affects fermion correlators in a boundary theory which has zero entropy at zero temperature. 
 Finally, in section \ref{sectionseven}, we discuss the relevance of our work to cuprate phase diagram and conclude with open questions as well as extensions for future work. In particular, we contemplate the extension of our results to superconducting backgrounds, which also do not possess a finite ground state degeneracy at zero temperature.

\section{Bulk Analysis }\label{sectiontwo}

As we alluded to above, we consider just one form of non-minimal coupling, in which a spin-$1/2$ fermion is coupled to  the gauge field through a dipole interaction\footnote{In even bulk dimensions, there is a similar interaction which includes a $\Gamma_5$. We will not consider this interaction in this paper.} of the form $F_{ab}\bar\psi \Gamma^{ab}\psi$, and propagates in the background of a Reissner-Nordstr\"om AdS$_{d+1}$  black-hole (hereafter, denoted by RN-AdS$_{d+1}$). Thus, we consider the bulk Lagrangian 
\begin{align}
\sqrt{-g}\,i\bar\psi (\slashed{D}-m- ip \slashed{F})\psi,
\end{align}
%in $d+1\geq4$ dimensions, 
in $d+1\geq4$ dimensions where
\begin{align} 
\bar\psi&=\psi\,\Gamma^{\underline t},\nonumber\\
\slashed{D}&=e_c^M\G^c\left(\partial_M+\frac{1}{4}\omega_M^{~ab}\G_{ab}-iqA_M\right),\nonumber\\
 \slashed{F}&=\frac{1}{2}\G^{ab} e_a^{M}e_b^{N}F_{MN}, 
 \end{align}
with $e^M_a$ and $\omega_M^{~ab}$ being the (inverse) vielbein and the spin connection, respectively. We denote the bulk coordinate indices by capital letters $M,N,\dots=\{t,x^i,r\}$  while the tangent space indices are denoted by $a,b,\cdots=\{\underline{t},  {\underline x}^i,\underline r\}$.  We will reserve the Greek indices $\mu, \nu, \ldots$ to denote boundary coordinate directions. We use Dirac matrices $\Gamma^{\underline t},\Gamma^{\underline 1},...,\Gamma^{\underline r}$ satisfying the Clifford algebra $\{\Gamma^a,\Gamma^b\}=2\eta^{ab}$. Also, $\G_{ab}=\frac{1}{2}[\Gamma_a, \Gamma_b]$. In what follows, we will rescale $p\to p L/(d-2)$ for convenience. 

%The significance of the parameters $m$ and $q$ are well-known: they correspond to the scaling dimension and global $U(1)$ charge of dual fermionic operators. The significance of $p$ in the dual field theory is not apparent however. In the holographic system, since $p$ is a coupling of the probe fermion to the bulk geometry, we expect that it may modify the correlation functions of the dual fermionic operator.
%We will argue below that this is a parameter that behaves much like doping in condensed matter systems. 

The RN-AdS$_{d+1}$ background has a metric and a gauge connection which can be written
\begin{align}
ds^2&=\frac{r^2}{L^2}\left[- f(r) dt^2 + d\vec x^2 \right]+\frac{L^2}{r^2} \frac{dr^2}{f(r)},\\
A&=\mu\left[1-(\frac{r_0}{r})^{d-2}\right]dt,
\end{align}
where
\begin{align}\label{fmu}
f(r)&=1-M\left(\frac{r_0}{r}\right)^d+Q^2\left(\frac{r_0}{r}\right)^{2(d-1)}\hskip-0.03in,\nonumber\\
\mu&=\left(\frac{d-1}{2d-4}\right)^{1/2}\,\frac{Qr_0}{L^2}, \qquad\qquad M=1+Q^2,
\end{align}
with $r_0$ being the horizon, given by the largest real root of $f(r_0)=0$. The temperature $T$ of this (black hole) background is given by
\bea
T=(d-2)\frac{r_0}{4\pi L^2}\left(\frac{d}{d-2}-Q^2\right).
\eea
From the above equation, one notes that the RN-AdS$_{d+1}$ black hole is extremal when $Q^2=d/(d-2)$ while the density and entropy remain finite. Since the background is invariant under $A_t\to -A_t$,  without loss of generality, we can choose $\mu$, or equivalently $Q$, to be positive. Thus, we can take $0< Q \leq \sqrt{d/(d-2)}$, where the equality corresponds to extremality.

To analyze the Dirac equations of the bulk fermion, we find it  more convenient to go to momentum space by Fourier transforming $\psi(r,x^{\mu})\sim e^{ik.x} \psi(r,k^{\mu})$, where $k^\mu=(\omega, {\vec k})$. The Fourier transform of the Dirac operator $\slashed{D}$ is of the form
\begin{align}
\slashed{D}&=\frac{r}{L}\sqrt{f(r)}\,\Gamma^{\underline r}\left[\partial_r  +\frac{f'(r)}{4f(r)} +\frac{d}{2r}\right]\nonumber\\
&\,-i\frac{L}{r\sqrt{f(r)}} \Gamma^{\underline t}\left[\omega+qA_t(r)\right]+i\frac{L}{r}\vec k\cdot\vec\Gamma,
\end{align}
while
\begin{align}
\slashed{F}=(d-2)\frac{\mu}{r_0}\left(\frac{r_0}{r}\right)^{d-1} \Gamma^{{\underline r}{\underline t}}.
\end{align}
To decouple the Dirac equations, we introduce projectors $\Gamma_\pm=\frac{1}{2}(1\pm \Gamma^{\underline r}\Gamma^{\underline t}\hat k\cdot\vec\Gamma)$ and write $\psi_\pm(r)=r^{d/2} f(r)^{1/4}\Gamma_\pm\psi(r)$. Without loss of generality, we set $k_1=k$ and $k_{i\neq 1}=0$, and take the basis 
\begin{align}\label{GammaBasis}
 \Gamma^{\underline r} &= \begin{pmatrix}
   -\sigma_3 \otimes\mathds{1} & 0 \\
    0&   -\sigma_3 \otimes\mathds{1}
   \end{pmatrix}, \hskip 0.03in\Gamma^{\underline t} = \begin{pmatrix}
   i\sigma_1 \otimes\mathds{1} & 0 \\
    0&   i\sigma_1 \otimes\mathds{1}
   \end{pmatrix},\nonumber\\ \Gamma^{\underline 1} &= \begin{pmatrix}
   -\sigma_2 \otimes\mathds{1} & 0 \\
    0&   \sigma_2 \otimes\mathds{1}
   \end{pmatrix}.
\end{align}
where $\sigma_j$'s are the Pauli matrices, and $\mathds{1}$ is a $2^{\frac{d-3}{2}}$-dimensional identity matrix for odd values of $d$, and $2^{\frac{d-4}{2}}$-dimensional for $d$ even. Note that by choosing $k_1=k$ and $k_{i\neq 1}=0$  the rest of the gamma matrices do not appear in the Dirac equations. So, we did not bother to include those in \eqref{GammaBasis}.
One then finds
%The equation of motion then falls apart into two decoupled equations
%\begin{align}\label{DiracEqInRN}
%\left(\frac{r^2}{L^2}\sqrt{f(r)}\,\partial_r+\frac{r}{L}m\,\sigma_3\right)\psi_{\pm}&=
%\frac{i\sigma_2}{\sqrt{f(r)}}\left[\omega
%  +\mu q
%  \left(1-\frac{r_0^{d-2}}{r^{d-2}}\right)\right]\psi_{\pm}\nonumber\\
%  &-\left(\mu p\frac{r_0^{d-2}}{r^{d-2}}\pm k\right) \sigma_1\psi_{\pm}.
%\end{align}
%
\begin{align}\label{DiracEqInRN}
\hskip -0.08in\frac{r^2}{L^2}\sqrt{f(r)}\,\partial_r\psi_{\pm}&=\frac{i\sigma_2}{\sqrt{f(r)}}\left[\omega
  +\mu q
  \left(1-\frac{r_0^{d-2}}{r^{d-2}}\right)\right]\psi_{\pm}\nonumber\\
&\hskip -0.08in-\sigma_1\left(\mu p\frac{r_0^{d-2}}{r^{d-2}}\pm k\right) \psi_{\pm}-\sigma_3\frac{r}{L}m\psi_{\pm}.
\end{align}
%\begin{widetext}\bea\label{DiracEqInRN}
%\left(\frac{r^2}{L^2}\sqrt{f(r)}\,\partial_r+\frac{r}{L}m\,\sigma_3\right)\psi_{\pm}=
%\left[\frac{i\sigma_2}{\sqrt{f(r)}}\left[\omega
%  +\mu q
%  \left(1-\frac{r_0^{d-2}}{r^{d-2}}\right)\right]-\left(\mu p\frac{r_0^{d-2}}{r^{d-2}}\pm k\right) \sigma_1\right]\psi_{\pm}.
%\eea
%\end{widetext}
We see that the Pauli coupling modifies the appearance of $k$ in the above Dirac equations. %As we will explain below, this feature can lead to a gap in the spectral density.
To see the effects of $p$ more clearly, consider the solutions of the Dirac equations \eqref{DiracEqInRN} in the asymptotic and near horizon regimes. Asymptotically, the solutions behave as 
%\begin{align}
%\psi_\pm(u,\omega,k) \sim u^{-d/2}\left[u^{mL} {\rm a}_\pm(\omega,k)(1+...)\begin{pmatrix}
%0 \\ 1 \end{pmatrix}+u^{-mL} {\rm b}_\pm(\omega,k)(1+...)\begin{pmatrix}
%1 \\ 0 \end{pmatrix}\right] .
%\end{align}
\begin{align}\label{psipmAsymptotics}
\psi_{\pm}(r, \omega, k)&= a_\pm (\omega,k)\,\,r^{mL} \, \begin{pmatrix}
0 \\ 1 \end{pmatrix}\Big[1+\cdots\Big]\nonumber\\
&+b_\pm  (\omega,k)\,r^{-mL} \begin{pmatrix}
1 \\ 0 \end{pmatrix}\Big[1+\cdots\Big].
\end{align}
%\bea\label{psipmAsymptotics}
%\psi_{\pm}(r, \omega, k)&=& a_\pm (\omega,k)r^{mL} \Big[1+\cdots\Big]\begin{pmatrix}
%0 \\ 1 \end{pmatrix}\nonumber\\
%&+&b_\pm  (\omega,k) r^{-mL} \Big[1+\cdots\Big]\begin{pmatrix}
%1 \\ 0 \end{pmatrix}.
%\eea
The effect of $p$ asymptotically is to modify the subleading terms. For $m\in [0,\frac{1}{2})$ both terms in \eqref{psipmAsymptotics} are normalizable and one can choose either $a_\pm$ or $b_{\pm}$ to be the sources for the dual fermion operator in the boundary theory. In this paper, we take $m\in [0,\frac{1}{2})$ and consider the conventional quantization where $a_{\pm}$ are the sources. Thus, the dual fermion operator has dimension $\Delta=\frac32+m$. 
%In this paper, we mostly consider the $m=0$ case, \ie $\Delta =3/2$. 
Choosing in-falling boundary conditions near the horizon results in a retarded correlator of the form 
\bea\label{ConventionalGreenFunctions}
G_{R}(\omega, k) = \begin{pmatrix}
   G_{+}(\omega, k)\, \mathds{1} & 0 \\
    0&      G_{-}(\omega, k)\, \mathds{1}
   \end{pmatrix},
\eea
with $G_{\pm}(\omega, k) = b_{\pm}(\omega, k)/a_{\pm}(\omega, k)$. Note that the Dirac equations \eqref{DiracEqInRN} imply $G_+(\omega,k)=G_-(\omega,-k)$. 

%\begin{figure*}
% \begin{center}
%\includegraphics[width=68mm]{ImGMP0Left.jpg}
%\hskip1.2in
%\includegraphics[width=67mm]{ImGMP45Right.jpg}
%\vskip-0.1in\caption{\footnotesize{${\rm Im}\,G_{-}(\omega,k)$ for $p=0$ and  $p=4.5$. A gap is clearly visible around $\omega=0$ in the second plot. We show just $G_-(\omega,k)$ to avoid cluttering the plots. $G_+(\omega,k)$ can be recovered using the relation $G_+(\omega,k)=G_-(\omega,-k)$. We have set $L=r_0=1$.} }
%\end{center}
%\end{figure*}
%In this note, we consider the extremal case (zero temperature). 
When the background is extremal, $f(r)$ has a double zero at the
horizon, $f(r)\sim d(d-1)(1-r_0/r)^2+\cdots$, and this fact makes taking
the limit of $ \omega \to 0$ of the equations \eqref{DiracEqInRN} near the horizon subtle. To take care of the subtlety, one realizes \cite{Faulkner:2009wj,Edalati:2009bi} that near the horizon (in which the geometry approaches $\rm{AdS}_2\times \mathds{R}^{d-1}$ for $T=0$) the equations for $\psi_{\pm}$ in \eqref{DiracEqInRN} organize themselves as functions of $\zeta = \omega L_2^2/(r-r_0)$ with $L_2= L/\sqrt{d(d-1)}$ being the radius of AdS$_2$. 
%as functions of $\zeta$, where
%%$\equiv \omega\eta$, where 
%\begin{align}\label{inner}
%r-r_0=\frac{\omega L_2^2}{\zeta}, \qquad\qquad L_2=\frac{L}{\sqrt{d(d-1)}},
%\end{align}
%with $L_2$ being the radius of AdS$_2$. 
The coordinate $\zeta$ is the suitable radial coordinate for the $\ads_2$ part of the near horizon region, and in this region, we can write $\psi_{\pm}$ in terms of $\zeta$ and expand in powers of  $\omega$ as follows
\begin{align}\label{innerPhi}
\psi_{I\pm}(\zeta)&=\psi^{(0)}_{I\pm}(\zeta)+\omega\,\psi^{(1)}_{I\pm}(\zeta)+\omega^2 \psi^{(2)}_{I\pm}(\zeta)+\cdots.
\end{align}
%we match the inner and outer region solutions from which we can analytically determine the fermion retarded Green functions at low frequencies. Basically, we impose infalling boundary condition at the horizon in the $\zeta$ coordinate, and match the inner and outer expansions in the so-called ``matching region" where the  $\zeta\to 0$ and $\omega/\zeta\to 0$ limits are taken.  Since the Dirac equations \eqref{DiracEqInRN} are linear,  and we also require that the solutions for the higher order terms in the inner and outer region expansions do not include terms proportional to  the zeroth-order solutions near the matching region, we just need to match $\psi^{(0)}_{I\pm}(\zeta)$  to $\psi^{(0)}_{O\pm}(r)$ near that region. 
%Near the horizon (the AdS$_2$ region), we have in the extremal case
%\begin{align}
%f(u)=(u-1)^2+...,\qquad v_\pm(u)\sim \frac{\omega}{\sqrt{6}(u-1)}+\frac{\frac53\omega+q\mu\pm\sqrt{6}p\mu}{\sqrt{6}}+...] 
%\end{align}
%
%and so for $\omega\neq 0$, we change coordinates to $u=1+\frac{\omega}{\mu\zeta}$ \ack{fix} so that
%\[
%v_\pm(\zeta)=\frac{\omega}{\sqrt{f(\zeta)}}\left(1+\frac{q}{\zeta+\omega/\mu}\right)\pm \frac{p\mu\zeta}{\zeta+\omega/\mu}
%\]
%The coordinate $\zeta$ is the suitable radial coordinate for the AdS$_2$ part of the near horizon region, and in this region, we can write $\psi_{\pm}$ in terms of $\zeta$ and expand in powers of  $\omega$ as follows
%\begin{align}\label{innerPhi}
%\psi_{I\pm}(\zeta)&=\psi^{(0)}_{I\pm}(\zeta)+\omega\,\psi^{(1)}_{I\pm}(\zeta)+\omega^2 \psi^{(2)}_{I\pm}(\zeta)+\cdots.
%\end{align}
Now, substituting \eqref{innerPhi} into \eqref{DiracEqInRN}, we find that to leading order
%\begin{align}\label{leadinglinner}
%-\psi^{(0)\prime\prime}_{I\pm}(\zeta)=i\sigma_2\left(1+\frac{q e_d}{\zeta}\right)-\frac{L_2}{\zeta}\left[m\sigma_3+\left(p e_d\pm\frac{kL}{r_0}\right)\sigma_1\right]\psi^{(0)}_{I\pm}(\zeta),
%\end{align}
\begin{align}\label{leadinglinner}
\psi^{(0)\prime}_{I\pm}(\zeta)&=\frac{L_2}{\zeta}\left[m\sigma_3+\left(c_d\frac{p}{L} \pm\frac{kL}{r_0}\right)\sigma_1\right]\psi^{(0)}_{I\pm}(\zeta)\nonumber\\
&-i\,\sigma_2\left(1+\frac{q e_d}{\zeta}\right)\psi^{(0)}_{I\pm}(\zeta),
\end{align}
where $e_d=1/\sqrt{2d(d-1)}$, and $c_d=1/\left[(2d-4)e_d\right]$. Equations \eqref{leadinglinner} are identical to the equations of motion for massive spinor fields \cite{Faulkner:2009wj} with masses $(m, \tilde m_+)$ and $(m,\tilde m_-)$ in AdS$_2$, where $\tilde m_\pm$ are time-reversal violating mass terms, with the identification
\begin{align}
\tilde m_\pm =c_d\frac{p}{L} \pm\frac{kL}{r_0}.
\end{align}
Thus,  $\psi^{(0)}_{I\pm}(\zeta)$ are dual to spinor operators ${\cal O}_{\pm}$ in the IR CFT with conformal dimensions $\delta_{\pm}=\nu_k^{\pm}+\frac{1}{2}$ where 
\begin{align}\label{nupm}
\nu_k^{\pm}&=\sqrt{m_{k\pm}^2L_2^2-q^2e_d^2-i\epsilon}, \nonumber\\
m_{k\pm}^2\hskip -0.05in &=m^2 +\left(c_d\frac{p}{L}\pm\frac{kL}{r_0}\right)^2.
\end{align}
We see that turning on $p$ modifies the scaling in the infrared in an important way -- effectively, the momentum is pushed up and down by $p$. We will explore the details of this in what follows.

One can write a formal expression for the fermion retarded correlator \eqref{ConventionalGreenFunctions} at low frequency in terms of the retarded Green functions  of the IR CFT spinor operators ${\cal O}_{\pm}$.  As shown in \cite{Faulkner:2009wj}, such a formal expression is extremely useful in analyzing the small $\omega$ behavior of the boundary theory Green functions. This is done by matching the inner AdS$_2$ and outer AdS$_4$ solutions in the so-called ``matching region" where the  $\zeta\to 0$ and $\omega/\zeta\to 0$ limits are taken. In so doing, one finds that the coefficients $a_\pm(\omega,k)$ and $b_\pm(\omega,k)$ in \eqref{ConventionalGreenFunctions} are given by
\begin{align}\label{xpm}
a_\pm(\omega,k)&=\left[a^{(0)}_\pm+\omega\,a^{(1)}_\pm+{\cal O}\left(\omega^2\right)\right]\nonumber\\
&\,+ \left[ {\tilde a}^{(0)}_\pm+\omega\, {\tilde a}^{(1)}_\pm+{\cal O}\left(\omega^2\right)\right] {\cal G}^\pm_k(\omega),\\ 
b_\pm(\omega,k)&=\left[b^{(0)}_\pm+\omega \,b^{(1)}_\pm+{\cal O}\left(\omega^2\right)\right]\nonumber\\ 
&\,+ \left[ {\tilde b}^{(0)}_\pm+\omega \,{\tilde b}^{(1)}_\pm+{\cal O}\left(\omega^2\right)\right] {\cal G}_k^\pm(\omega),\label{ypm}
\end{align}
where $a^{(n)}_\pm, {\tilde a}^{(n)}_\pm, b^{(n)}_\pm$ and  ${\tilde b}^{(n)}_\pm$  are all functions of $k$ and can, in principle, be determined numerically. Also, in the above expressions, ${\cal G}_k^{\pm}(\omega)$ denote the retarded Green functions of the dual IR CFT operators ${\cal O}_{\pm}$ which are given by \cite{Faulkner:2009wj}
\begin{align}\label{tworetgreen}
{\cal G}_k^\pm(\omega)=c_\pm(k) \,\omega^{2\nu_\pm}, 
\end{align}
with
\begin{align}
&c_\pm(k)=e^{-i\pi\nu_\pm}\frac{\G(-2\nu_\pm)\G(1+\nu_\pm-iq e_d)}{\G(2\nu_\pm)\G(1-\nu_\pm-iq e_d)}\times\nonumber\\
&\hskip 0.9in \frac{(m+i\tilde m_\pm)L_2-iqe_d-\nu_{\pm}}{(m+i\tilde m_\pm)L_2-iqe_d+\nu_{\pm}}.
\end{align}
Note that the expressions \eqref{xpm} are \eqref{ypm} are not valid when $2\nu_\pm$ is an integer. In such cases there would be additional terms like $\omega^n {\rm log} \, \omega$ (with $n$ being a positive integer) on the right hand sides of  \eqref{xpm} and \eqref{ypm}.

In order to obtain $G_\pm(\omega, k)$ for generic values of $\omega$ and $k$, one must solve the Dirac equations \eqref{DiracEqInRN} numerically. For numerical purposes, it is convenient to work with dimensionless quantities.  So, we rescale $r$, $\omega$ and $k$ in the Dirac equations \eqref{DiracEqInRN} by defining
 \begin{align}
r\to r_0 u, \qquad \omega \to \frac{r_0}{L^2} \,\omega, \qquad k\to \frac{r_0}{L^2}\, k.
 \end{align}
%From now on, we work with the dimensionless quantities $u$, $\omega$ and $k$. 
It is also more convenient to convert the Dirac equations \eqref{DiracEqInRN} into the so-called flow equations \cite{Liu:2009dm, Iqbal:2008by}. For that, we first write $\psi_{\pm}^T=(\beta_{\pm}, \alpha_{\pm})$ and define $ \xi_{\pm}=\beta_{\pm}/\alpha_{\pm}$, in terms of which the Dirac equations \eqref{DiracEqInRN} then reduce to the non-linear flow equations
%\begin{align}
%\psi_{\pm}=\begin{pmatrix}
%\beta_{\pm} \\  \alpha_{\pm} \end{pmatrix}, \qquad \qquad \xi_{\pm}=\frac{\beta_{\pm}}{\alpha_{\pm}}.
%\end{align}
%$\psi_\pm=\begin{pmatrix} y_\pm\cr z_\pm\end{pmatrix}$ and $\xi_\pm=\frac{y_\pm}{z_\pm}$. 
%The equations \eqref{DiracEqInRN} then reduce to flow equations
\bea\label{FlowEquations}
\hskip-0.3in u^2\sqrt{f(u)}\partial_u\xi_{\pm}&=&-2(mL)u\,\xi_{\pm}\nonumber\\
&&+\left[v_{-}(u)\mp k\right]+\left[v_{+}(u)\pm k\right]\xi^2_{\pm}, \hskip 0.06in
\eea
%\begin{align}
%u^2\sqrt{f(u)}\pa_u \xi_\pm+2mL u\xi_\pm +\left[\left(v_-(u)\mp k\right)+\left(v_+(u)\pm k\right)\xi_\pm^2\right]=0\label{decoupledeom}
%\end{align}
where
\begin{align}\label{vpm}
v_{\pm}(u)=\frac{1}{\sqrt{f(u)}}\left[\omega+Q q \left(1-u^{2-d}\right)\right]\pm Q p \,u^{2-d}.
\end{align}
To obtain the retarded Green functions of the boundary theory operators, one has to choose the infalling boundary condition at the horizon for the (dual) bulk fields \cite{Son:2002sd, Iqbal:2009fd}. Expressed in terms of $\xi_{\pm}$, the infalling boundary condition for $\psi_{\pm}^T=(\beta_{\pm}, \alpha_{\pm})$ at the horizon  translates into 
\begin{align}
\hskip -0.09in\xi_\pm (u=1)=\left\{ \begin{array}{ll}
 i & \hskip 0.03in \omega \neq 0,   \\
 (mL_2 -\nu_{\pm})/(qe_d+ \tilde m_{\pm}L_2)&   \hskip 0.03in \omega =0. 
\end{array}
\right.
\end{align}
The matrix of Green functions \eqref{ConventionalGreenFunctions} then takes the form
\begin{align}\label{XiPmGreenFunctions}
G_R(\omega, k) = \lim_{\epsilon\to0}\,\epsilon^{-2mL}\begin{pmatrix}
   \xi_{+}\, \mathds{1} & 0 \\
    0&       \xi_{-}\, \mathds{1}
   \end{pmatrix}\Big|_{u=\frac{1}{\epsilon}},
\end{align}
where one picks the finite terms as $\epsilon\to 0$.  Up to normalization, the fermion spectral function is defined by 
\begin{align}
A(\omega, k) \equiv {\rm Tr}\, {\rm Im}\, G_R(\omega,k).
\end{align}

\section{Continuum and Bound States}
\label{sectionthree}

In this section, we will study the effects of small and negative values of $p$. As we will see, in this regime, there is a (non-Fermi-liquid) Fermi peak whose properties change as we vary $p$. In this regime, there is some similarity to the properties of Fermi surfaces studied in \cite{Liu:2009dm, Faulkner:2009wj}. As we change parameters, the scaling dimensions change, and we can pass from a non-Fermi liquid (NFL) to a marginal Fermi liquid (MFL) and on to fermions which have some resemblance to Landau Fermi liquids (FL). However, we will see clearly that there is a positive value of $p$  beyond which the Fermi peak ceases to exist. In what follows, we will mostly set $m=0$ and $q=1$ (as we vary $p$), although similar results hold for a range of these parameters. 

\begin{figure}
\centering
\includegraphics[width=60mm]{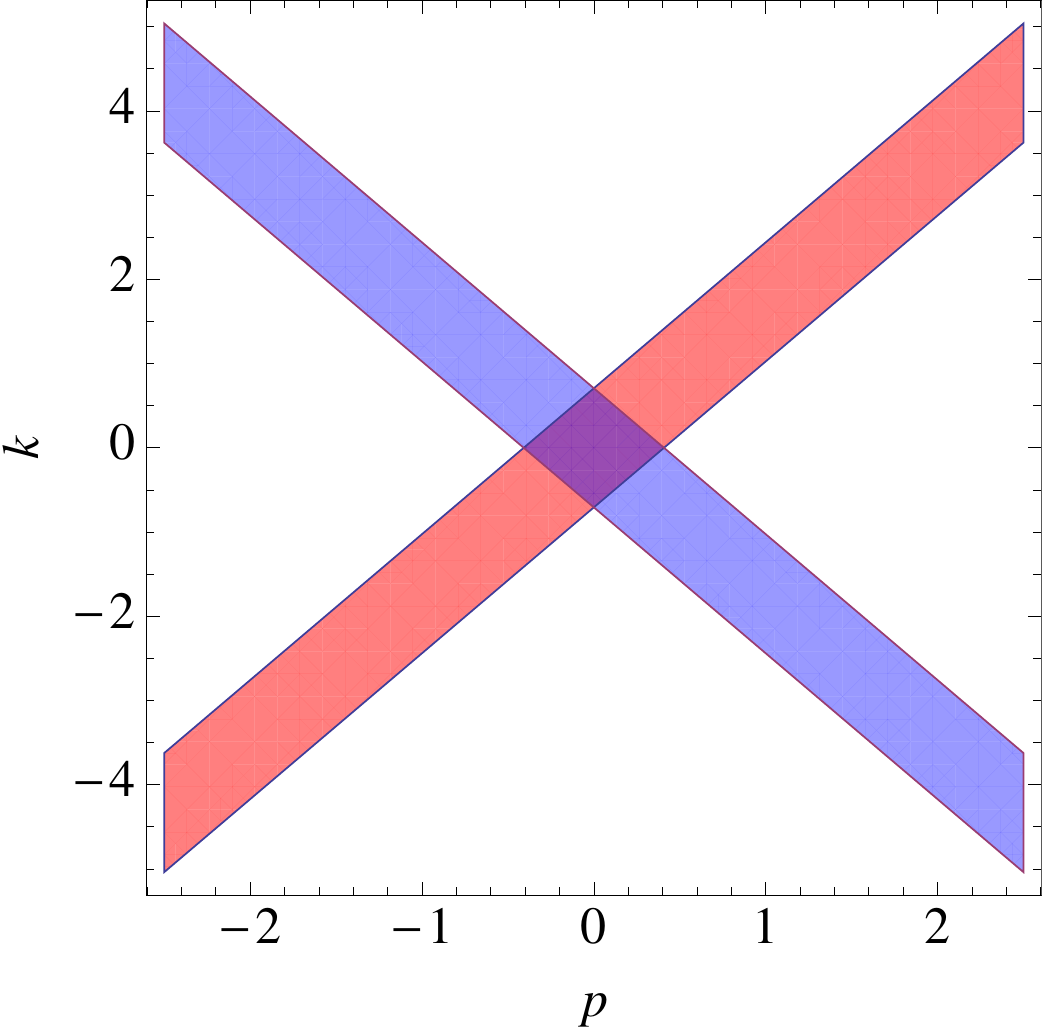} 
\caption{\label{Ipmvsp}\footnotesize{Plots of $\cal I_\pm$ versus $p$ (for $d=3$, $\Delta_\psi=3/2$ and $q=1$).  The red band depicting $\cal I_-$ is where ${\rm Im}\, G_{-} (\omega,k)$ becomes oscillatory at small $\omega$. The blue band ($\cal I_+$) shows the region where ${\rm Im}\, G_{+} (\omega,k)$ is oscillatory (at small $\omega$).} }
\end{figure}

\begin{figure*}
\centering
\hskip -0.12in \includegraphics[width=57mm]{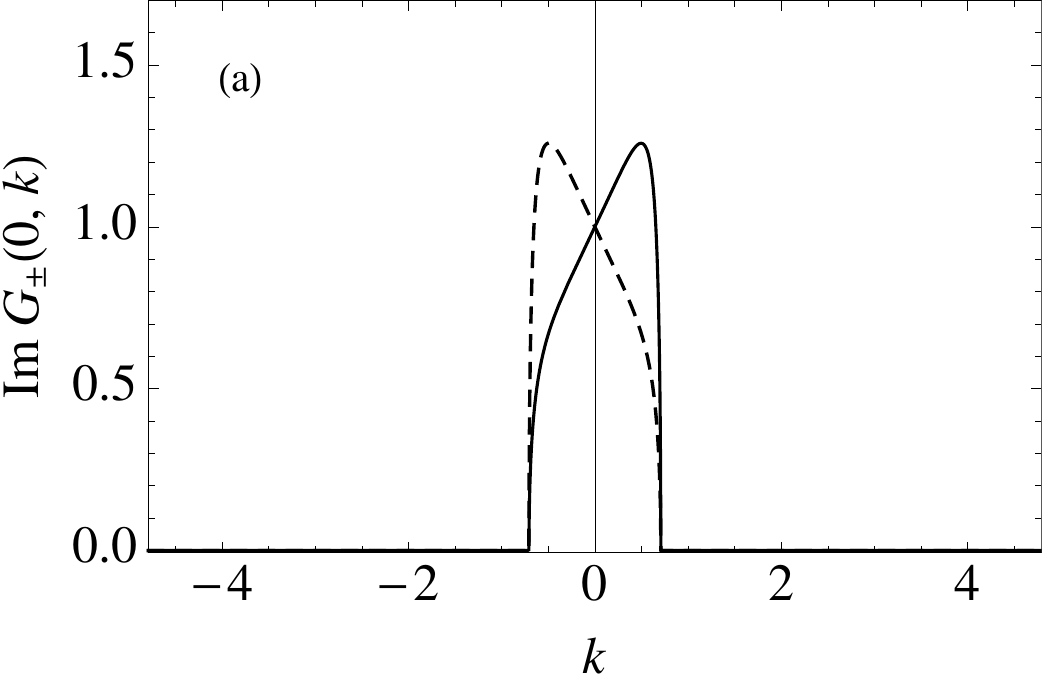}
\hskip 0.1in
\includegraphics[width=57mm]{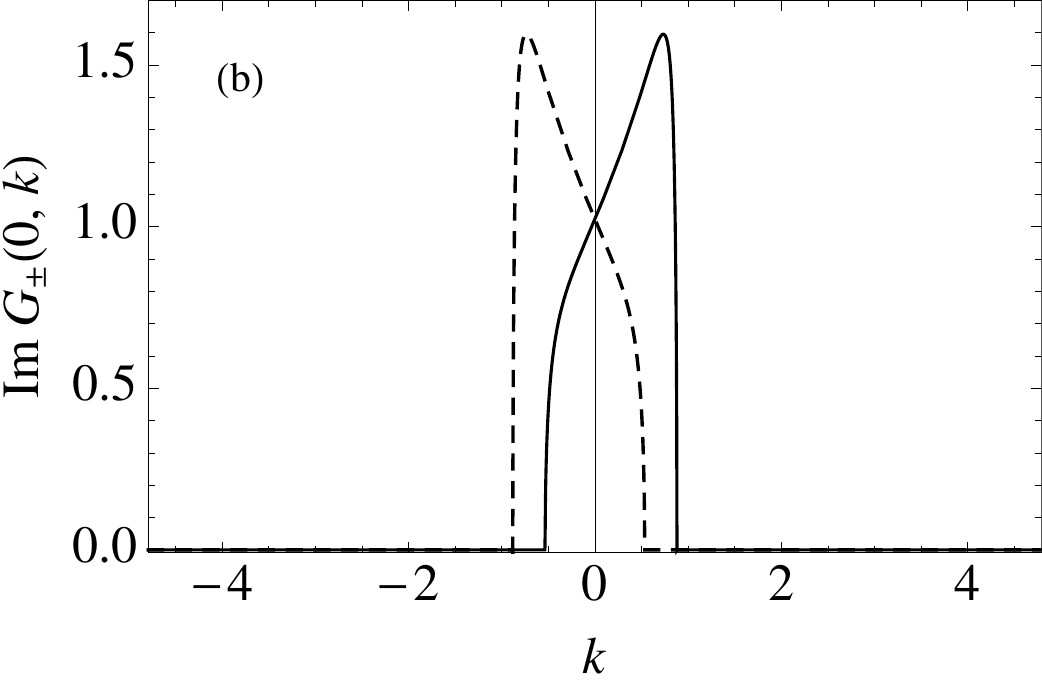}
\hskip 0.1in
\includegraphics[width=57mm]{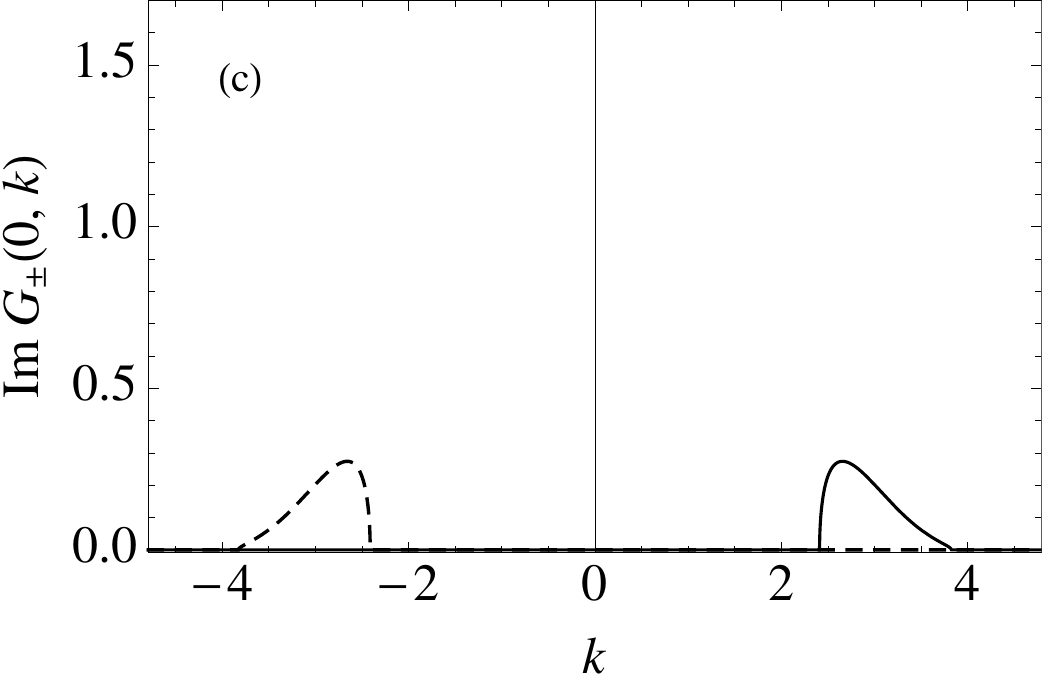}
\vskip -0.1in
\caption{\label{ImGPMOzerovsK}\footnotesize{Plots of ${\rm Im}\, G_{-} (0,k)$ (solid line) and ${\rm Im}\, G_{+} (0,k)$ (dashed line) for (a) $p=0$, (b) $p=0.1$ and (c) $p=1.8$. We set $d=3$, $\Delta_\psi=3/2$ and $q=1$. Similar plots can be obtained for  negative values of $p$ by switching the solid lines with the dashed lines. Focusing on positive $p$, we see that the maximum value of ${\rm Im}\, G_{\pm} (0,k)$ increases as $p\to 1/\sqrt{6}$, after which (namely, for $p> 1/\sqrt{6}$) it rapidly decreases.} }
\end{figure*}

We begin by focussing on the regime that has been called log-oscillatory in \cite{Liu:2009dm, Faulkner:2009wj}, in which Fermi peaks do not occur. There are some important changes when $p\neq 0$ that we will explain below.
When $q^2>2m^2L^2$, there exists a range of momenta $k\in \cal I_+$
for which the dimension of the IR CFT operator ${\cal O}_+$ becomes
imaginary. Similarly,  for  $k\in \cal I_-$ the dimension of ${\cal
  O}_-$ becomes imaginary.  Here,  we have defined ${\cal I}_\pm =
(\mp c_d\,p -k_o, \mp c_d\,p +k_o) $ with $k_o=
\sqrt{(q^2/2)-m^2L^2}$.   (Figure \ref{Ipmvsp} shows plots of $\cal
I_\pm$ versus $p$.) Consequently, ${\rm Im}\, G_{\pm} (0,k)$ is
generically non-vanishing for $k\in \cal I_\pm$, respectively; see
Figure \ref{ImGPMOzerovsK} for plots of ${\rm Im}\, G_{\pm} (0,k)$ as a function of $p$.
At $p=0$, one has $\cal I_+ = \cal I_-\equiv {\cal I}$. This case was analyzed in \cite{Liu:2009dm, Faulkner:2009wj} where it was found that for $k\in {\cal I}$ and for small $\omega$,  both ${\rm Im}\, G_{\pm} (\omega,k)$ are periodic in ${\rm log} \,\omega$, with the same period. 

At $p=0$, the range of momenta for which ${\rm Im}\, G_{\pm} (\omega,k)$ become log-oscillatory at small $\omega$ is the same for each, namely $k\in {\cal I}$. This degeneracy does not persist for non-zero $p$, hence the fermion spectral function $A(\omega,k)$ will also have non-oscillatory components. For $p\in [-k_o/c_d,k_o/c_d]- \{0\}$, both ${\rm Im}\, G_{\pm} (\omega,k)$ show log-oscillatory behavior (with different periods, though) only for $k \in {\cal I}_+ \cap {\cal I}_-$. For $|p|>k_o/c_d$ where  ${\cal I}_+ \cap {\cal I}_- = \emptyset$, one finds that in the regime where ${\rm Im}\, G_{-} (\omega,k)$ shows log-oscillatory behavior, ${\rm Im}\, G_+ (\omega,k)$ is not oscillatory and vice versa. 

For real $\nu_k^\pm$, the boundary conditions for $\xi_\pm(u=1)$ at $\omega=0$ are real. Since the equations \eqref{FlowEquations} are real, one deduces that ${\rm Im}\, G_{\pm} (0,k)=0$. Thus, ${\rm Re}\, G_{\pm} (0,k)= G_{\pm} (0,k)=b_\pm^{(0)} /a_\pm^{(0)}$. ${\rm Re}\, G_{\pm} (0,k)$ may have poles which would be given generically by the zeros of $a_\pm^{(0)}$. Each zero of $a_\pm^{(0)}$ defines a Fermi momentum $k_{\rm F}$, given that $b_\pm^{(0)}$ do not vanish as $k\to k_{\rm F}$. Since $G_-(\omega,k)=G_+(\omega,-k)$, vanishing of $a_{-}^{(0)}$  at some $k=k_{\rm F}$ implies that $a_{+}^{(0)}$ vanishes at  $k=-k_{\rm F}$. So, in order to find $k_{\rm F}$, we can just analyze the zeros of $a_{-}^{(0)}$. From the asymptotic behavior of $\psi_-$ which is given in $\eqref{psipmAsymptotics}$, together with \eqref{xpm} and the definition of $\psi_{-}^T=(\beta_{-}, \alpha_{-})$, it is easy to see that,  at $\omega=0$, $\alpha_-(u,k) = a_-^{(0)}\,u^{mL} + \cdots$ as $u\to \infty$. (Also, note that at $\omega=0$, $\beta_-(u,k) = b_-^{(0)}\,u^{-mL} + \cdots$ as $u\to \infty$.)
%\begin{align}
%\alpha_-(u,k) = a_-^{(0)}\,u^{mL} + \cdots \,\,\,\,{\rm as}\,\,\,\, u\to \infty.
%\end{align}
So the $k_{\rm F}$'s define a set of momenta for which, at $\omega=0$,
$\psi_{-}(u,k)$ becomes normalizable (a `bound state') as $u\to
\infty$. To find the $k_{\rm F}$'s, we analyze the equation for $\alpha_-(u,k)$ as follows.
%or equivalently, the zeros of $\alpha_{-}(u=\infty, \omega=0,k)$. In other words, 

Plugging $\psi_{-}^T=(\beta_{-}, \alpha_{-})$ into the Dirac equations \eqref{DiracEqInRN}, one obtains a set of two coupled linear differential equations for $\alpha_{-}$ and $\beta_{-}$. Setting $\omega=0$ and decoupling these two equations, we obtain
%the equation for $\alpha_{-}(u,k)$ is found to be
\begin{align}\label{alphamOmegaZero}
-\frac{u^2\sqrt{f(u)}}{v^{0}_{-}(u)+ k} \, \partial_u\left(\frac{u^2\sqrt{f(u)}}{v^{0}_{+}(u)- k} \,\partial_u\right)\alpha_{-} =\alpha_-\, , \\
-\frac{u^2\sqrt{f(u)}}{v^{0}_{+}(u)- k} \, \partial_u\left(\frac{u^2\sqrt{f(u)}}{v^{0}_{-}(u)+ k} \,\partial_u\right)\beta_{-} =\beta_-\, , \label{betamOmegaZero}
\end{align}
where we have set $m=0$ for convenience. In \eqref{alphamOmegaZero}, the superscript ``0" on $v_{\pm}(u)$ indicates that we have set $\omega=0$ in \eqref{vpm}.  Once again, we set $d=3$ and $q=1$ in what follows. As we vary $p$,  we look (numerically) for the momenta  $k_{\rm F}$ for which  $\alpha_{-}(u=\infty, k_{\rm F})=0$, given an appropriate boundary condition for $\alpha_{-}(u, \omega=0,k)$ at the horizon. Indeed, solving \eqref{alphamOmegaZero} near the horizon, one easily obtains that  $\alpha_{-}(u,k)\sim f(u)^{\pm \nu_k^{-}/2}$ as $u\to 1$. Because by assumption we are in a regime where $\nu_k^{-}$ is real and positive\footnote{For $\nu^-_k=0$,  one finds that $\alpha_{-}(u\to1,k)=a\, (1+\cdots)+ b\,{\rm log}(u-1)(1+\cdots)$ where the dots represent terms which vanish as $u\to 1$, and $a$ and $b$ are some constants. In order for $\alpha_{-}(u,k)$ not to blow up at the horizon, one should then choose $b=0$.} (and, in fact, generically irrational), $f(u)^{- \nu_k^{-}/2}$ blows up as $u\to 1$. Thus, $\alpha_{-}(u\to 1,k)\sim f(u)^{\nu_k^{-}/2}$  is the regular horizon boundary condition that should be chosen. 

\begin{figure}
\centering
\hskip -0.4in \includegraphics[width=60mm]{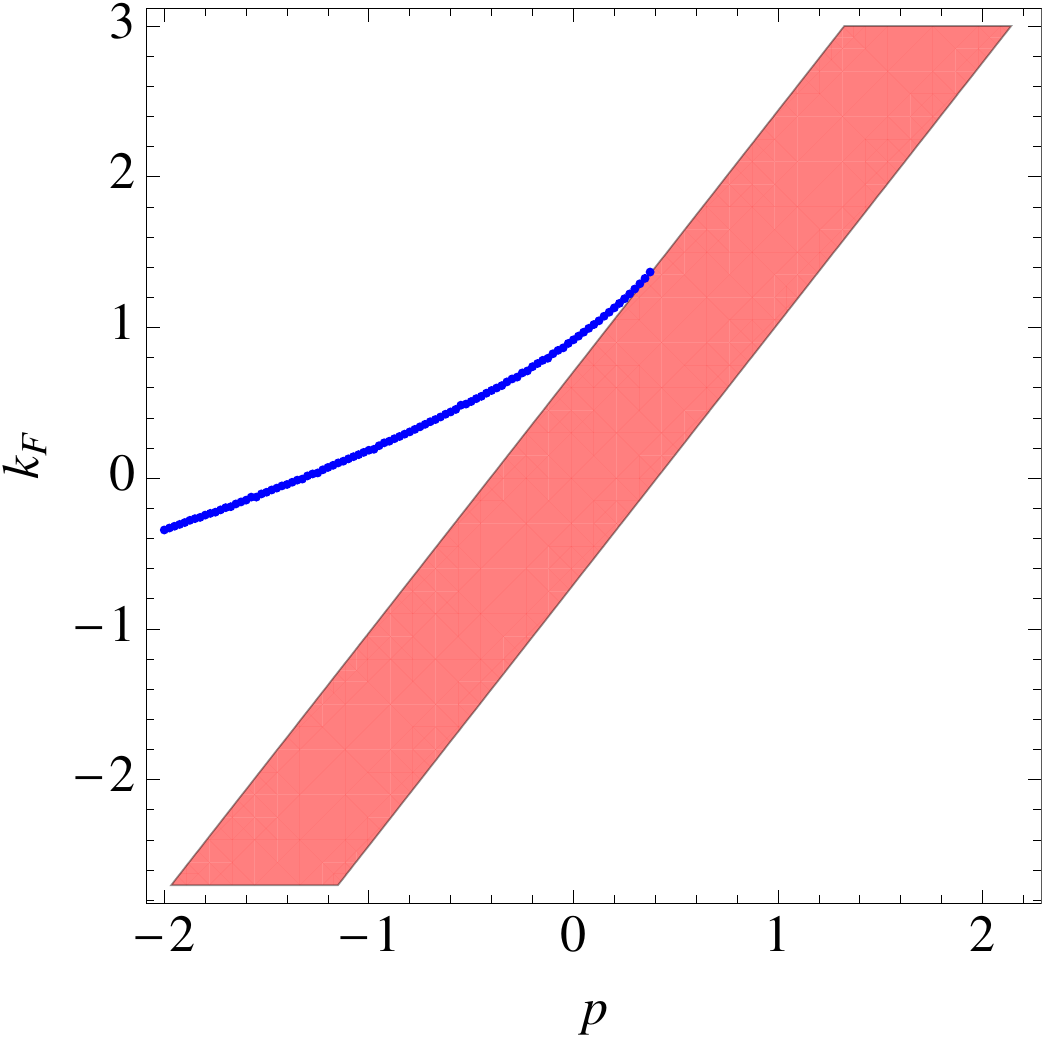} 
\caption{\label{kFvsp}\footnotesize{$k_{\rm F}$'s (shown by blue dots) versus $p$. For $p\leq 1/\sqrt{6}$ there is a single Fermi surface for each $p$. For $p> 1/\sqrt{6}$ we do not find Fermi surfaces. The orange band shows the oscillatory region ${\cal I}_-$.} }
\end{figure}

\begin{figure*}
\centering
\hskip -0.2in \includegraphics[width=65mm]{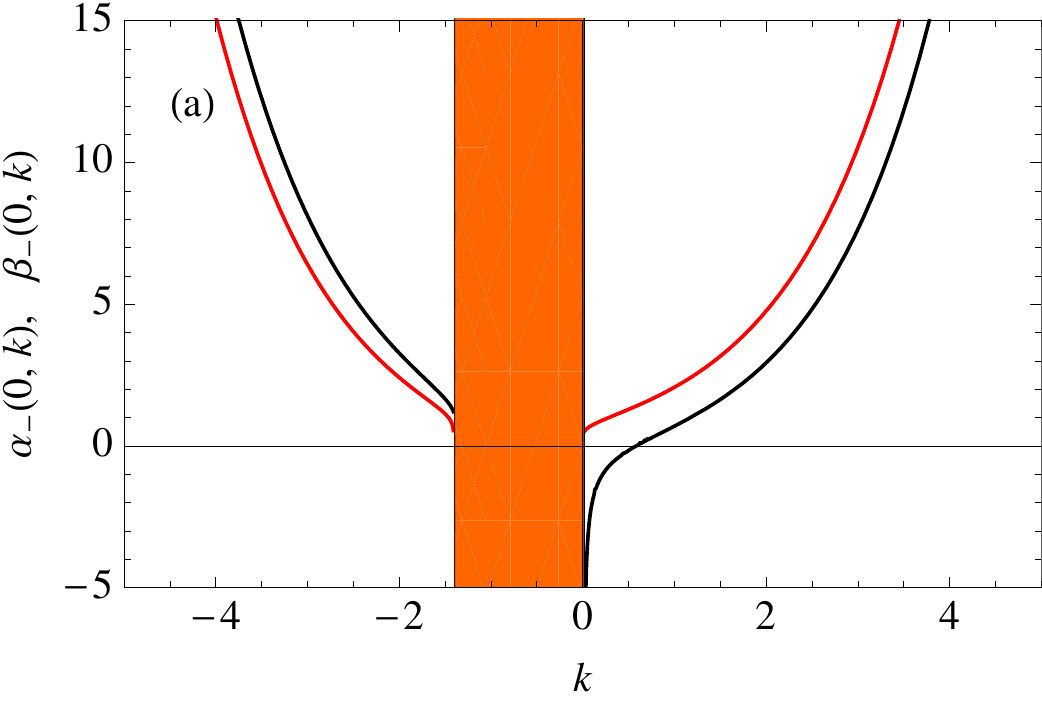} \qquad\qquad
\includegraphics[width=65mm]{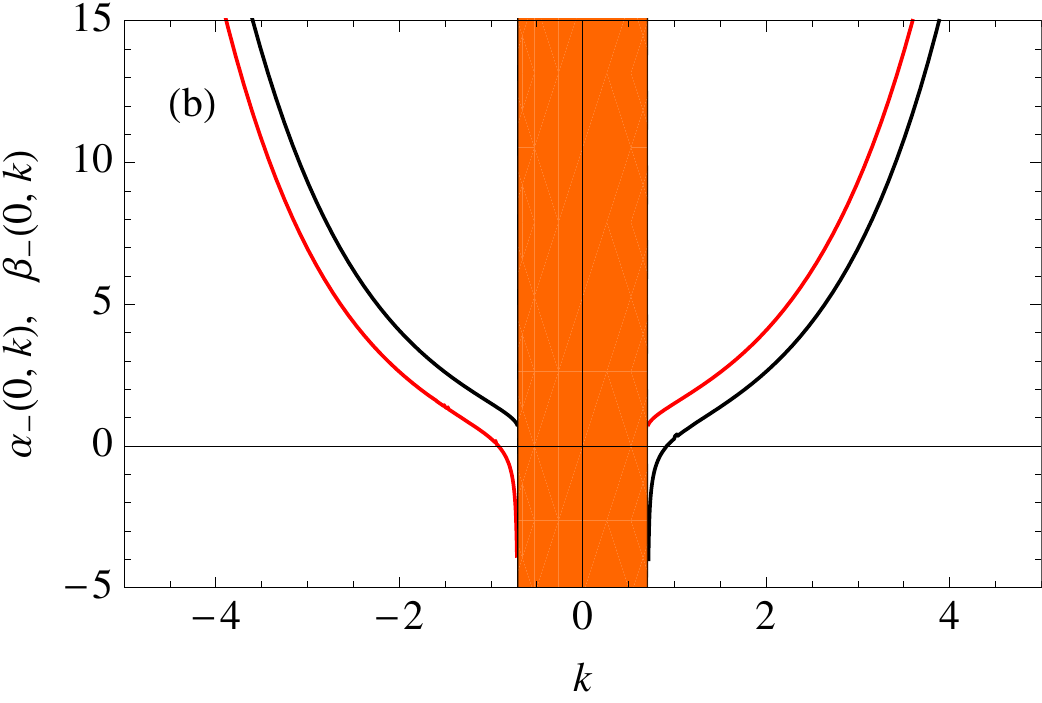} \\
\hskip -0.2in\includegraphics[width=65mm]{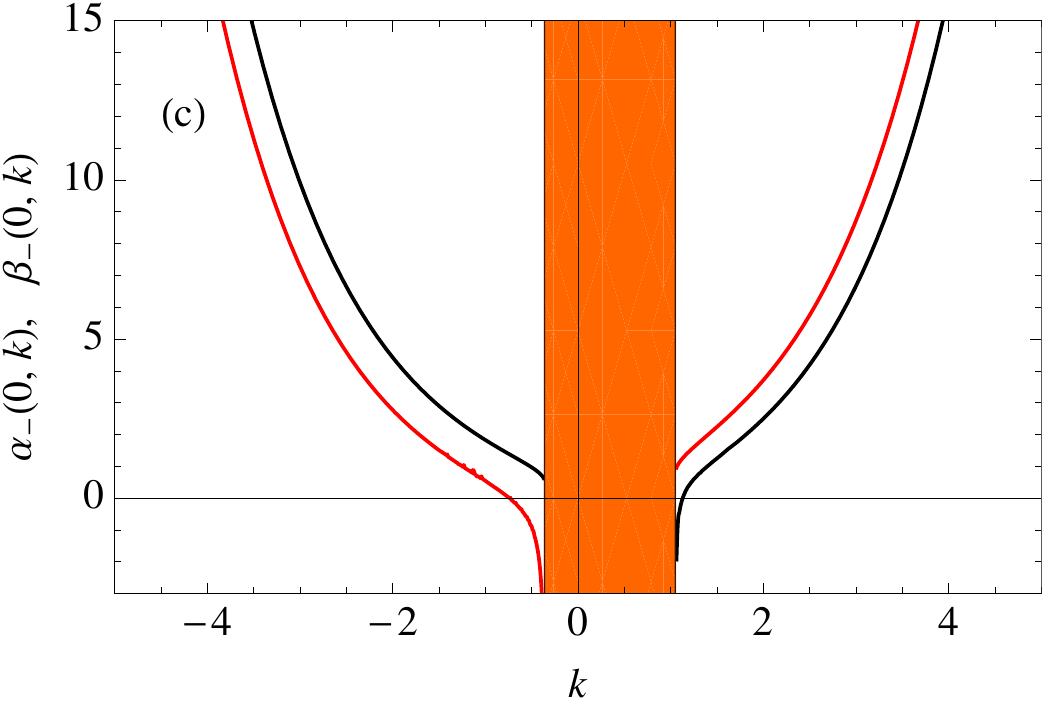} \qquad\qquad
\includegraphics[width=65mm]{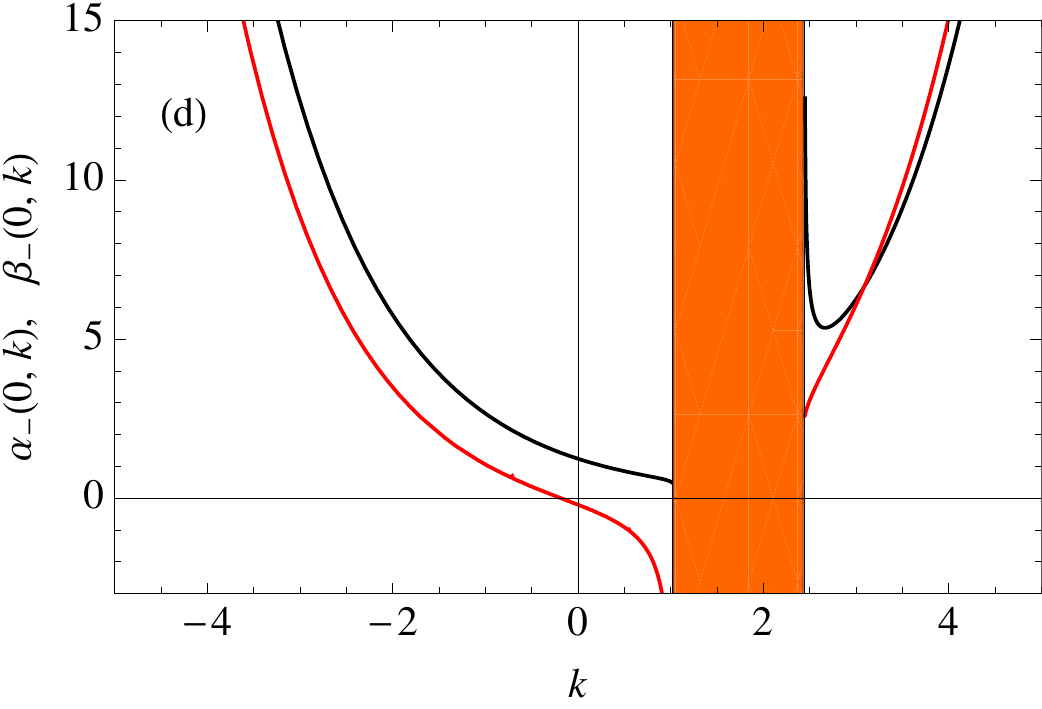} 
\caption{\label{aMinusvsK}\footnotesize{Plots of $\alpha_{-}(u=\infty, k)$ (black curves) and $\beta_{-}(u=\infty, k)$ (red curves) versus $k$ for (a) $p=-0.4$, (b) $p=0$, (c) $p=0.2$, (d) $p=1$. The orange strip in each plot shows the oscillatory region ${\cal I}_-$. The plots are generated for $d=3$, $m=0$ and $q=1$. By $k\to -k$, similar plots could be obtained for $\alpha_{+}(u=\infty, k)$ and $\beta_{+}(u=\infty, k)$.} }
\end{figure*}

In Figure \ref{kFvsp} we have plotted such values of $k_{\rm F}$ as a function of $p$. Starting with negative values of $p$ (while keeping $q=1$ fixed), $k_{\rm F}$ increases as we raise $p$ causing it to move towards the boundary of the oscillatory region $\cal I_-$.  As $p$ approaches $1/\sqrt{6}$ from below, $k_{\rm F}$ approaches $\sqrt{2}$ (in units of $r_0/L^2$, from below). The blue dots in Figure \ref{kFvsp} show the location of $k_{\rm F}$'s versus $p$ and the red band depicts the oscillatory region. There is a single Fermi surface for each $p$ as we increase $p$ up to $p=1/\sqrt{6}$. We have explicitly checked that $\beta_- (u=\infty ,k=k_{\rm F})$ does not vanish, so $k=k_{\rm F}$ are genuine poles of ${\rm Re}\,G_{-} (0,k)$. At $p=1/\sqrt{6}$, $k_{\rm F}=\sqrt{2}$, and as a result $\nu^-_{k_{\rm F}}$ vanishes (recall that $d=3$, $m=0$ and $q=1$). (At this point, and in fact at any point in which $2\nu^\pm_{k_{\rm F}}\in\mathbb{Z}$, the analysis should be more carefully done, as logarithms must be included.)
For $p>1/\sqrt{6}$, we do not see a Fermi surface as
$\alpha_{-}(u=\infty, k)$ does not vanish outside the oscillatory
region. We have checked this numerically up to $p=10$,   and, given
the observed behavior of  $\alpha_{-}(u=\infty, k)$, we do not expect it to change as we increase $p$ further. Indeed, Figure \ref{aMinusvsK} shows plots of $\alpha_{-}(u=\infty, k)$ versus $k$ for sample values of $p$. We have also plotted $\beta_{-}(u=\infty, k)$, shown by the red curves in Figure \ref{aMinusvsK}. 
%It is clear from these plots that $\beta_{-}(u=\infty, k_{\rm F})\neq 0$. 

Following \cite{Faulkner:2009wj}, the excitations around these Fermi surfaces can be analyzed. Using \eqref{xpm} and \eqref{ypm}, near $k=k_{\rm F}$ and at small $\omega$, $G_{-}(\omega, k)$ takes the form 
\begin{align}\label{SmallOmegaGM}
\hskip -0.01 in G_{-}(\omega, k)\approx\frac{b_{-}^{(0)}(k_{\rm F})}{\partial_k a_{-}^{(0)}(k_{\rm F})k_{\perp}+\omega\,a^{(1)}_{-}(k_{\rm F})+{\tilde a}^{(0)}_{-}(k_{\rm F}) {\cal G}_{k_{\rm F}}^{-}(\omega)},
\end{align}
where $k_\perp =k-k_{\rm F}$.  Suppose the denominator in \eqref{SmallOmegaGM} vanishes at some  $\omega_{*}(k)= {\rm Re\,} \omega_{*}(k) - i\, {\rm Im\,} \omega_{*}(k)$. For $p=0$, the dispersion relation, the width and the residue of the pole were worked out in detail in \cite{Faulkner:2009wj}. Parts of the data in these quantities (such as the scaling of the dispersion relation) come from the IR CFT (or, equivalently, the AdS$_2$ part of the near horizon geometry), and other parts (such as Fermi velocity) from the UV physics. For non-zero $p$, the data which come from the IR CFT will be slightly modified according to \eqref{nupm} whereas the data coming from the UV physics could be substantially modified. 

\newcommand{\z}{{\, {\rm z}}}

For  $-0.53 < p<1/\sqrt{6}$, we find that $1/2>\nu^-_{k_{\rm F}}>0$. As a result, the small $\omega$ excitations around $k=k_{\rm F}$ will have  a (non-Fermi liquid) dispersion relation ${\rm Re\,} \omega_{*}(k)\propto k_{\perp}^z$ and  a width ${\rm Im\,} \omega_{*}(k)\propto k_{\perp}^z$ where $z=1/(2\nu^{-}_{k_{\rm F}})$. Thus, for this range of $p$, ${\rm Im\,} \omega_{*}(k)/{\rm Re\,} \omega_{*}(k)$ does not vanish as $\omega\to 0$, implying that these excitations are not stable. Note that the residue at the pole is given by $Z\propto k_\perp^{z-1}$. At $p=-0.53$, $\nu^-_{k_{\rm F}}=1/2$ and the excitations near the Fermi
surface are of the marginal Fermi liquid type. For $-1.54<p<-0.53$,
$1>\nu^-_{k_{\rm F}}>1/2$, hence, the small $\omega$ excitations around
$k=k_{\rm F}$ will have  a linear dispersion relation ${\rm Re\,}
\omega_{*}(k)\propto k_{\perp}$ and  a width ${\rm Im\,}
\omega_{*}(k)\propto k_{\perp}^{2\nu^-_{k_{\rm F}}}$. So, for this range of
$p$, these excitations are stable as ${\rm Im\,} \omega_{*}(k)/{\rm
  Re\,} \omega_{*}(k)\to 0$ as $\omega\to 0$. It is in this sense that
we refer to this region as the Fermi liquid.  Also, we found that
$k_{\rm F}$ goes through zero at $p=-1.317$, signifying that the excitations
change over from `particle-like' to `hole-like'. Consequently, we find
that simply by varying $p$, we can tune from a Fermi liquid, $p<-0.53$
to a marginal Fermi liquid at $p=-0.53$ to a generic
non-Fermi liquid for $-0.53<p<1/\sqrt{6}$ and finally to a Mott
insulator for $p$ sufficiently large.  Precisely how the system
behaves for large values of $p$ will now be addressed in the
quasi-normal mode analysis.
%FL, $p<0.53
%One can easily consider more negative values of $p$ and do a similar
%analysis. For example, to the extent that we have checked (which is
%up to $p=-2$) we found that $\nu^-_{k_{\rm F}}\geq1$ for $p\leq-1.53$.
%Consequently, we find that simply by varying $p$, we can tune from a
%FL, $p<0.53
 
\section{quasi-normal Modes and Stability}\label{sectionfour}

In this section we analyze the poles of $G_{\pm} (\omega,k)$ and, in particular,  discuss how they move in the complex $\omega$-plane as  we vary $k$. Since $G_+(\omega,-k)=G_-(\omega,k)$,  without loss of generality, we can just focus on the poles of $G_-(\omega,k)$. We denote the poles of $G_-(\omega,k)$ by $\omega_{*}(k)$. At small 
$\omega$, the poles of $G_-(\omega,k)$ can be worked out
semi-analytically. Indeed, for $p=0$, the small $\omega$ poles were
worked out in detail in  \cite{Faulkner:2009wj} for $k\in {\cal I}$ as
well as $k$ close to $k_{\rm F}$ where it was argued that such poles are all
located in the lower half of the complex  $\omega$-plane. Since the
arguments of  \cite{Faulkner:2009wj} are mainly based on the data
coming from the IR region (the near horizon AdS$_2$ region of the
background), they can easily be extended to non-zero values of $p$,
where results similar to those in the case of $p=0$ are obtained. For
example, at small $\omega$ and for $k\in \cal I_{-}$, where ${\rm
  Im}\, G_{-} (\omega,k)$ is oscillatory, the poles (for a fixed $k$) are exponentially separated on a straight line which is in the lower half $\omega$-plane. (The line is diagonally oriented, \ie it ends on the $\omega=0$ branch point.) Equivalently, for those values of $p$ for which there exists a Fermi surface, the small $\omega$ poles near $k=k_{\rm F}$ are all located in the lower half $\omega$-plane. In particular,   if $\nu^-_{k_{\rm F}}<1/2$, then $\omega_{*}(k\to k_{\rm F})$ as a function of $k$ follows a straight line in the lower half of the complex $\omega$-plane. The semi-analytic arguments of  \cite{Faulkner:2009wj}, and their generalizations to non-zero $p$, are applicable only for small $\omega$, and for $k\in \cal I_\pm$, or  when $k$ is near $k_{\rm F}$. Nevertheless, on general grounds, one expects the poles to be located in the lower half $\omega$-plane beyond the small $\omega$ regime (and, of course, for all values of $k$). 
%This is because we do not expect a single fermion to condense.  
To find the poles beyond the small $\omega$ regime, one is usually forced to do numerics which, in the context of the AdS/CFT correspondence, involves performing some quasi-normal mode analyses in the bulk.

Besides isolated poles, $G_\pm(\omega,k)$ at zero temperature will have a branch cut (at least for small $\omega$) which could be understood without doing the numerics. Note that since ${\cal G}_k^{\pm}(\omega)\sim \omega^{2\nu_k^\pm}$ appears in the expressions for  $a_{\pm}(\omega, k)$ and $b_{\pm}(\omega, k)$ in \eqref{xpm} and \eqref{ypm},  $G_\pm(\omega,k)$ will have a branch point at $\omega=0$ for generic values of $k$ (where $2 \nu_k^\pm$ are irrational), and a branch cut, which we take to be extended in the negative imaginary axis. For those values of $k$ for which $2\nu_k^\pm\in \mathbb{Z}$, there is still a branch  cut which is due to the appearance of logarithmic terms of the form $\omega^n {\rm log}\, \omega$ ($n\in \mathbb{Z}$) in the expressions for  $a_{\pm}(\omega, k)$ and $b_{\pm}(\omega, k)$. The branch cut seems to be a distinctive feature of the two-point retarded correlators of operators in the zero temperature $d$-dimensional boundary theory dual to the extremal RN-AdS$_{d+1}$ background. Indeed, the branch cut was observed explicitly in the correlators of scalar and spinor operators in \cite{Faulkner:2009wj, Denef:2009yy} as well as the conserved currents in the shear and sound channels in \cite{Edalati:2010hk, Edalati:2010pn}. As we will see below, this branch cut appears in our quasi-normal mode analysis. At finite temperature, however, the branch cut dissolves into a series of isolated poles on the negative imaginary axis. 

Generically, $G_R(\omega, k)$  will have poles  whenever $a_{\pm}(\omega,k)=0$. 
%So, equivalently,  we analyze whether the frequencies for which $a_\pm(\omega, k) =0$ can have positive imaginary part.  
In the context of the AdS/CFT correspondence, this problem could be addressed by computing the quasi-normal modes of $\psi_\pm$ in the RN-AdS$_{d+1}$ background, which are solutions to the Dirac equations \eqref{DiracEqInRN}  subject to the boundary conditions that they are infalling at the horizon and normalizable asymptotically. Except in very special cases, the generic values of the quasi-normal frequencies are usually computed numerically. We use the so-called Leaver's method \cite{Leaver:1990zz} for this purpose.
%, and analyze both the zero- and finite-temperature cases. 
For concreteness,  we take the boundary theory to be (2+1)-dimensional, \ie $d=3$. Our analysis can straightforwardly be extended to larger values of $d$. Also, as in the previous discussions, we consider  $m=0$ and $q=1$.

Substituting $\psi_{\pm}^T=(\beta_{\pm}, \alpha_{\pm})$ in the Dirac equations \eqref{DiracEqInRN}, and setting $m=0$, one finds 
\begin{align}\label{betapm}
u^2\sqrt{f(u)}\,\partial_u\beta_{\pm}&=\left[v_{-}(u)\mp k\right]\alpha_{\pm}\,,\\
u^2\sqrt{f(u)}\,\partial_u\alpha_{\pm}&=-\left[v_{+}(u)\pm k\right]\beta_{\pm}\,.\label{alphapm}
\end{align}
The equations for $\alpha_{\pm}$ are the relevant equations for obtaining the quasi-normal frequencies of $\psi_{\pm}$. Squaring the above equations, the decoupled  equations for $\alpha_{\pm}$ are easily obtained
\begin{align}\label{alphapm}
\frac{u^2\sqrt{f(u)}}{v_{-}(u)\mp k} \, \partial_u\left(\frac{u^2\sqrt{f(u)}}{v_{+}(u)\pm k} \,\partial_u\right)\alpha_{\pm} =-\alpha_\pm.
\end{align}
As we alluded to above, without loss of generality, we can focus on the quasi-normal frequencies of $\psi_-$ and just analyze the equation for $\alpha_{-}$ in \eqref{alphapm}. 
%This is because from the  Dirac equations \eqref{DiracEqInRN} one deduces that the equation for $\psi_+$ can be obtained from that of $\psi_-$ by $k\to -k$ (or, equivalently, $G_+(\omega,k)=G_-(\omega,-k)$ in the language of the boundary theory). 
%So, in what follows, we just analyze the equation for $\alpha_{-}$ in \eqref{alphapm}.
%Defining $\psi_{-}^T=(\beta_{-}, \alpha_{-})$, and setting $m=0$ in the Dirac equations \eqref{DiracEqInRN}, the equation for $\psi_+$ then yields two coupled equations for $\alpha_{-} $ and $\beta_{-}$,
%\begin{align}
%u^2\sqrt{f(u)}\,\partial_u\beta_{\pm}&=\left[v_{-}(u)\mp k\right]\alpha_{\pm}\,,\\
%u^2\sqrt{f(u)}\,\partial_u\alpha_{\pm}&=-\left[v_{+}(u)\pm k\right]\beta_{\pm}\,.
%\end{align}

In what follows, we switch to a new radial coordinate $z=1/u$ which is more convenient for doing the numerics in this section. In terms of the new radial coordinate,  the horizon is at $z=1$ and the asymptotic boundary at $z=0$.  The equation for $\alpha_{-}$ in \eqref{alphapm}  then becomes
\begin{align}\label{alpham}
\frac{\sqrt{f(z)}}{v_{-}(z)+ k} \, \partial_z\left(\frac{\sqrt{f(z)}}{v_{+}(z)- k} \,\partial_z\right)\alpha_{-} =-\alpha_{-}.
\end{align}

%\subsection{Zero Temperature}

To compute the quasi-normal modes of $\psi_{-}$, the behavior of $\alpha_{-}$ should be infalling at the horizon and normalizable at the boundary. As mentioned above, we use Leaver's method \cite{Leaver:1990zz} to compute the quasi-normal frequencies. For that, we first pull out the leading behavior of $\alpha_{-}$ at the horizon as well as the boundary and write  
\begin{align}\label{LeadingBehavioralpham}
\alpha_{-}(z)=e^{i\frac{\omega}{6(1-z)}}f(z)^{-i\left(\frac{\omega}{9}+\frac{q}{4\sqrt{3}}\right)} z\,\tilde \alpha_{-}(z).
\end{align}
Note that $\tilde\alpha_{-}(z=1)$ is a constant which could be set equal to unity as the equation for $\alpha_{-}$ is homogeneous.  %\sum_{m=0}^M a^{\pm}_m(\wn,\qn) \left(z-\frac{1}{2}\right)^m.
Next, we write $\tilde \alpha_{-}(z)$ as a power series in $z$ around a point $z_0=1/2$ (so that the radius of convergence of the series covers both the horizon and the boundary)
\begin{align}\label{tildealphamseries}
\tilde \alpha_{-}(z)=\sum_{m=0}^M \tilde \alpha^{-}_m(\omega,k) \left(z-\frac{1}{2}\right)^m.
\end{align}
Substituting \eqref{tildealphamseries} and \eqref{LeadingBehavioralpham} into \eqref{alpham},
%$M+1$ equations for $M+1$ unknowns $\tilde \alpha^{-}_m(\omega, k)$'s are obtained , and write them in a matrix form 
one obtains
\begin{align}
\sum_{m=0}^M A^{-}_{mp}(\omega,k) \, \tilde \alpha^{-}_m(\omega,k)=0,
\end{align}
where $A^{-}_{mp}(\omega,k)$ are the elements of a $(M+1)$ by $(M+1)$ matrix $A^{-}(\omega, k)$. The quasi-normal frequencies (for a fixed $k$) are then the solutions to 
\begin{align}\label{det}
\det~A^{-}(\omega,k)=0.
\end{align}
%In practice, one has to choose a value for $M$ such that  the solutions to \eqref{det}  converge. 
\begin{figure}
\centering
\includegraphics[width=70mm]{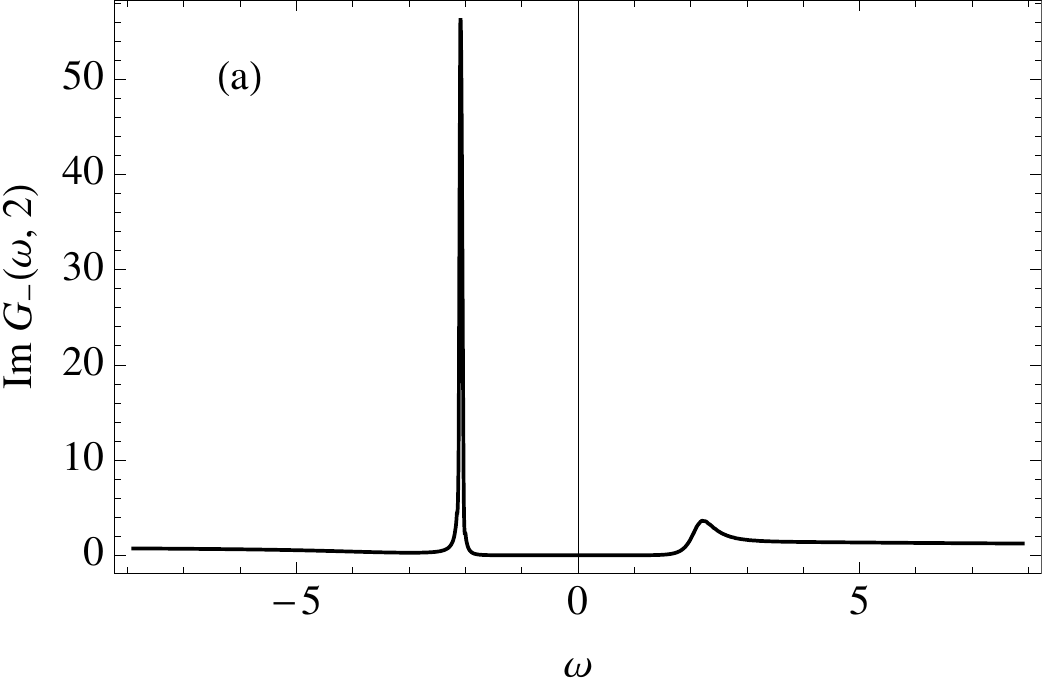} \\
\hskip -0.1in \includegraphics[width=72mm]{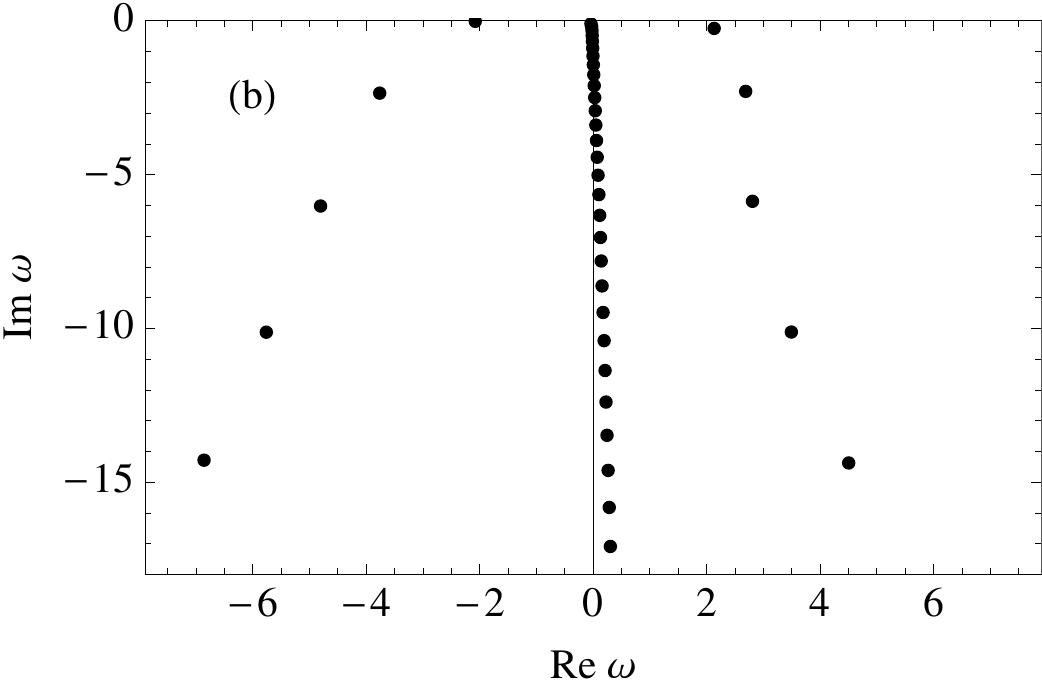} 
\caption{\label{ImGQNM}\footnotesize{(a) ${\rm Im}\,G_-(\omega, k)$ as a function of $\omega$ for $k=2$. (b) The quasi-normal frequencies of $\alpha_-$ for $k=2$. $d=3$, $p=5$, $q=1$ and  $m=0$ in both plots. Also, $M=250$.} }
\end{figure}
\begin{figure*}
\centering
\hskip -0.2in \includegraphics[width=65mm]{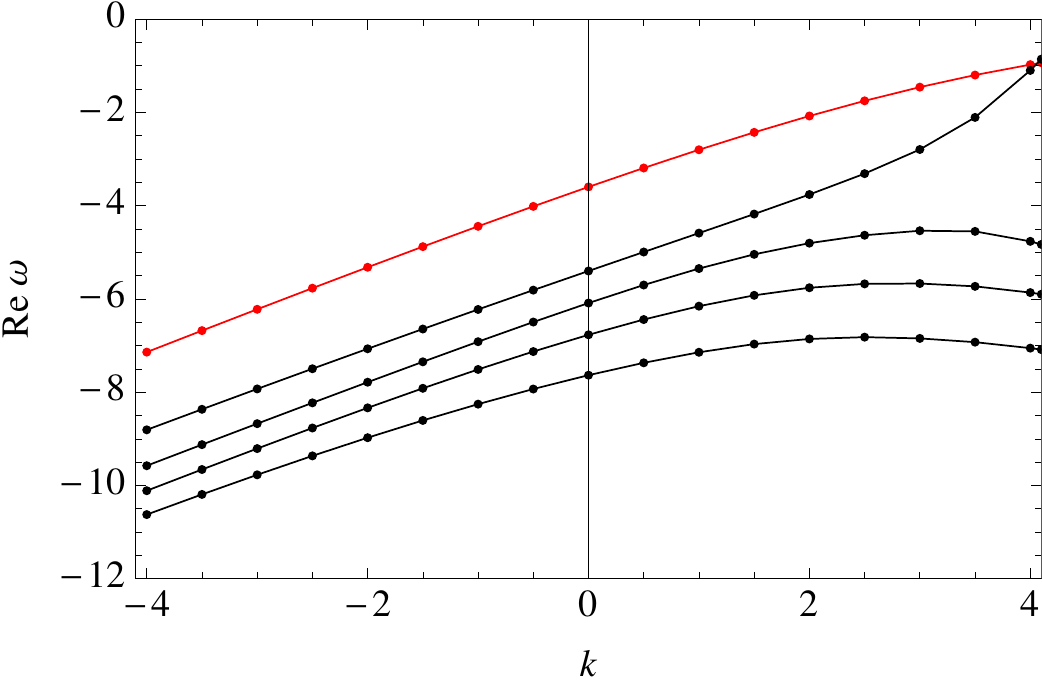} \qquad\qquad
\includegraphics[width=65mm]{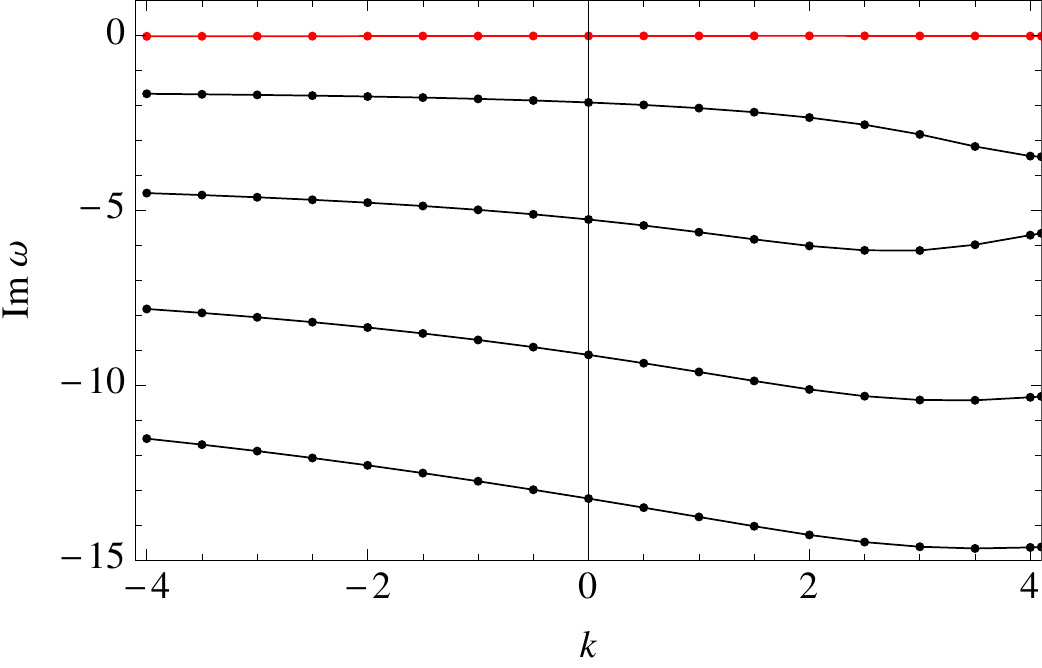} \\
\hskip -0.2in\includegraphics[width=65mm]{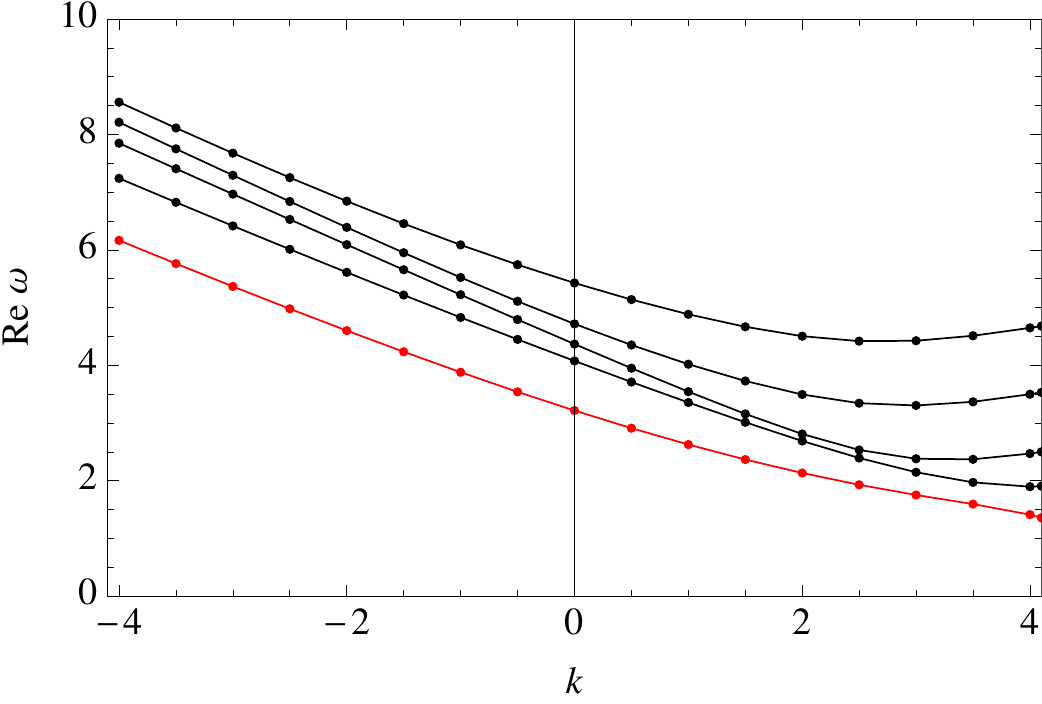} \qquad\qquad
\includegraphics[width=65mm]{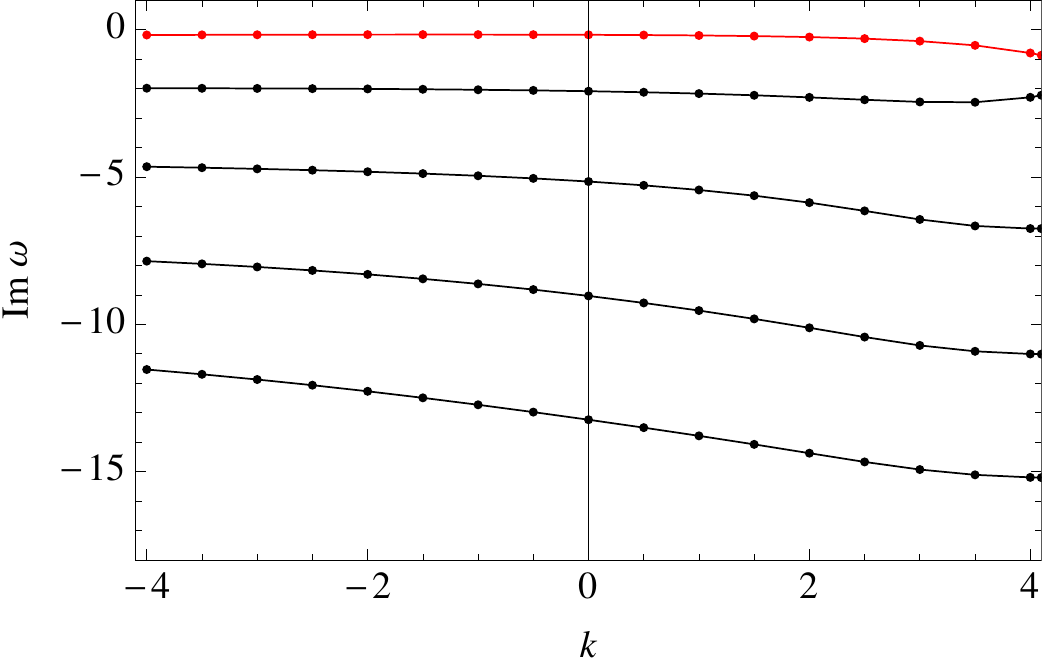} 
\caption{\label{DispersionofQNM}\footnotesize{The top two plots show the dependence on $k$ (dispersion relation) of the real and imaginary parts of the first five quasi-normal modes (depicted in Figure \ref{ImGQNM}(b)) on the left hand side of the negative imaginary axis.  The bottom two plots show the dispersion relation of the real and imaginary parts of the first five quasi-normal modes of Figure \ref{ImGQNM}(b) which are on the right hand side of the negative imaginary axis. The plots are generated for $d=3$, $p=5$, $m=0$ and $q=1$ and $M=250$. The red data corresponds to the mode closest to the real axis in the complex $\omega$-plane.} }
\end{figure*}

The bottom plot in Figure \ref{ImGQNM} shows the quasi-normal frequencies of $\alpha_-$ for $p=5$ and  $k=2$. To generate this plot we set $M = 250$.  Due to space limitations, the plot only shows a  handful of the quasi-normal frequencies. As $M$ is increased, the poles located along the negative imaginary axis become closer to one another, suggesting that their existence is due to taking $M$ to be finite, and in the limit of $M\to \infty$ they should indeed form the branch cut we mentioned earlier. On the other hand, increasing $M$ does not seem to change the qualitative behavior of the poles which are oriented almost diagonally on each side of the negative imaginary axis. Notice that the poles are all located on the lower half 
$\omega$-plane. As may be seen from the plot, the branch cut bends to the right for large negative values of ${\rm Im}\,\omega$. This behavior is different from the cases studied in \cite{Edalati:2010hk, Edalati:2010pn}  where the unbroken parity symmetry of the boundary theory forces the branch cut of the retarded correlators to stay on the negative imaginary axis. Indeed, the bending of the branch cut is similar to what was observed in \cite{Denef:2009yy} for the retarded correlators of charged scalar operators in the presence of a magnetic field. The top plot in Figure  \ref{ImGQNM} shows ${\rm Im}\,G_-(\omega,k=2)$ as a function of $\omega$. The location of the peak on the left hand side and the bump on the right hand side match quite well with the two quasi-normal frequencies which are closest to the real axis.  It is apparent that all of the other quasi-normal modes are relatively wide and individually have small residue.

The dispersion relation, $\omega_{*}(k)$, of the quasi-normal frequencies shown in Figure  \ref{ImGQNM}(b) can be computed numerically by following their motion in the complex $\omega$-plane as $k$ is varied. As shown in Figure \ref{DispersionofQNM}, the two quasi-normal frequencies which are closest to the real axis have markedly different dispersion relations than the rest (higher resonances).  The effects of these higher resonances become important as one goes beyond the small frequency and momentum approximation. For large $|k|$, 
%${\rm Im}\, \omega_{*}(k)$ 
the imaginary part of the quasi-normal frequencies is approximately constant (and negligible compared to the real part). In contrast, for large $|k|$, the real part  is proportional to $k$, with the constant of proportionality being equal to $\pm 1$ (within our numerical precision). This behavior is expected because at large frequency and momentum, the vacuum of the boundary theory is effectively Lorentz-invariant. So, the dispersion relation of the excitations should effectively be relativistic at large frequency and momentum where the effect of charge density is negligible. 

\begin{figure}
\centering
\includegraphics[width=75mm]{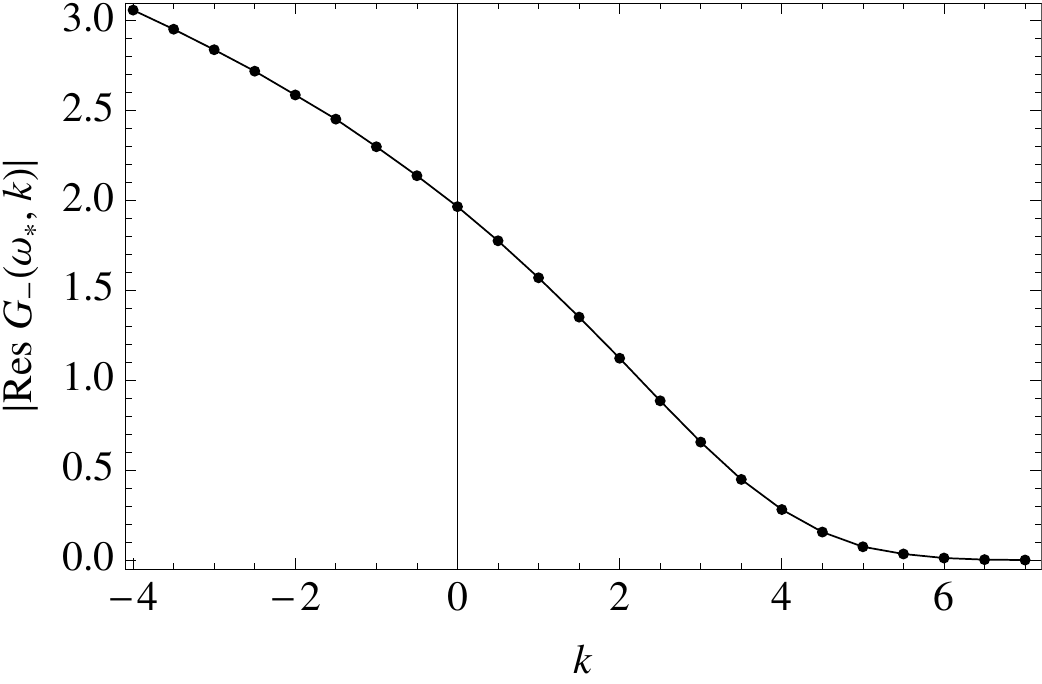} 
\caption{\label{ResGM}\footnotesize{ $|{\rm Res}\,G_- (\omega_*,k)|$ as a function of $k$ for the leading negative-frequency pole in Figure \ref{ImGQNM}(b) which is closest to the real axis and located to the left of the negative imaginary axis. We set $d=3$, $p=5$, $q=1$ and  $m=0$.} }
\end{figure}

\begin{figure}
\centering
\begin{tabular}{ccc}
\hskip -0.1in\includegraphics[width=42mm]{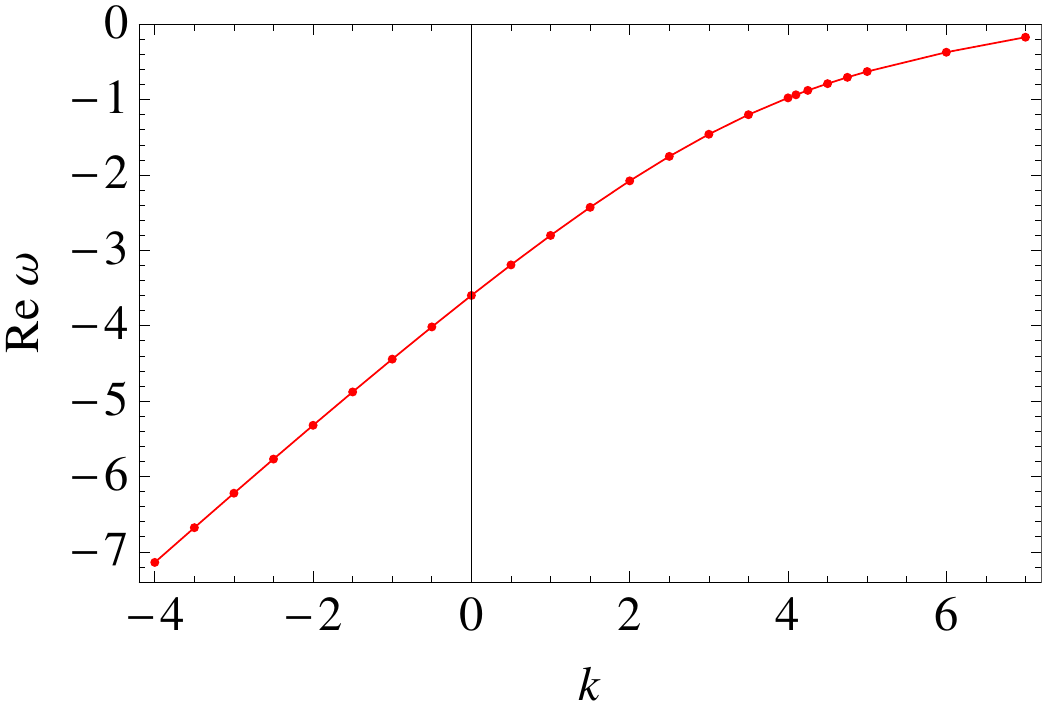} &
\includegraphics[width=45mm]{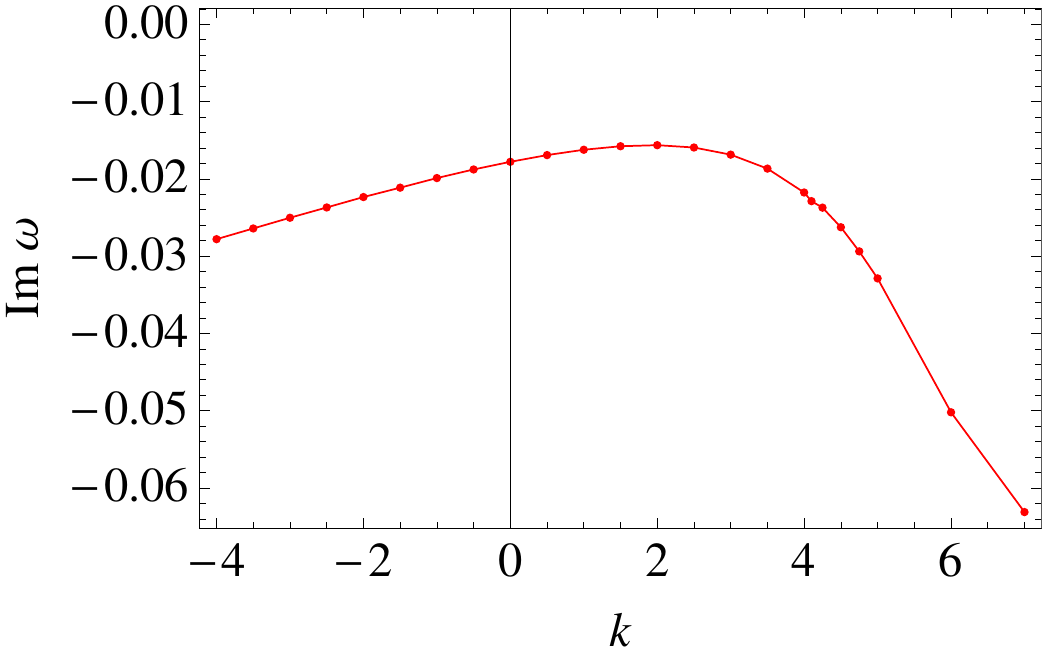} 
\end{tabular}
\caption{\footnotesize{A close-up of of the real (left plot) and imaginary (right plot) parts of the dispersion relation of the leading pole shown in Figure \ref{ImGQNM}(b) which is closest to the real axis and located to the left of the negative imaginary axis.} }
\label{LeadingLeftQNM}
\end{figure}

To have a better understanding of the spectrum, it is important to know how the residues (at the poles) behave as a function of $k$. In order to numerically compute the residues  of $G_- (\omega, k)$ at $\omega=\omega_*(k)$, we first developed series expansions for $\alpha_{-}(z; \omega, k)$ and $\beta_{-}(z; \omega, k)$ both near the horizon around $z=1-\epsilon $ and near the boundary around $z=\delta$. We then numerically integrated the (decoupled) differential equation  for $\alpha_{-}(z; \omega, k)$ and $\beta_{-}(z; \omega, k)$ from $z =1-\epsilon$ to $z=\delta$ and matched the numerically integrated solutions and their (first) derivatives with their boundary series expansions at $z = \delta$. In so doing, we were able to compute the residues  of $G_- (\omega, k)$ at $\omega=\omega_*(k)$  for a fixed $k$, denoted by ${\rm Res}\,G_- (\omega_*,k)$.  Repeating the same steps for different $k$'s, one can numerically obtain the dependence of the residues on $k$. We computed the $k$-dependence of the residues  of $G_- (\omega, k)$ for the leading negative-frequency pole (closest to the real axis). Shown in Figure \ref{ResGM} is the absolute value of ${\rm Res}\,G_- (\omega_*,k)$ as a function of $k$ for this pole. Also, the plots in Figure \ref{LeadingLeftQNM} show a close up of the real and imaginary parts of the dispersion relation of this pole. Comparing the plot of the residue to Figure \ref{ImGQNM}(a), it is now apparent why the gap forms: as the leading negative-frequency quasi-normal mode approaches $\omega=0$, its residue dies off quickly. Since no other mode has appreciable spectral weight, this accounts for the suppression of the spectral weight near $\omega=0$. For larger values of $p$, the residue falls off more rapidly, and consequently the gap widens.

\section{Finite Temperature}\label{sectionfive}
%\begin{figure*}
%\hskip -0.25in
%\begin{tabular}{ccc}
%\includegraphics[width=56mm]{SWP45.pdf} &
%\includegraphics[width=56mm]{SWQ165.pdf} &
%\includegraphics[width=56mm]{SWQ160.pdf} 
%\end{tabular}
%\caption{\footnotesize{A close-up of $A(\omega, k)$ as a function of
%$\omega$ for sample values of $k\in [0.1,5]$ for $p=4.5$. The left
%plot shows a gap of $\Delta/\mu\simeq0.0417$ at zero
%temperature. There is no gap for the middle ($T/\mu\simeq0.0133$) and
%the right ($T/\mu\simeq0.0218$ ) plots. As the temperature increases, more weight is seen around $\omega=0$. In particular, the middle plot indicates that the temperature for which the gap starts to close should be less than the gap $\Delta/\mu\simeq0.0417$.} }
%\end{figure*}

So far, our analysis has been at zero temperature.  However, there are
important aspects of Mott insulators that transpire at finite
temperature.  In particular, there are Mott insulators \cite{zylbert}
exhibiting a transition to a conducting state as the temperature
is increased.  The classic example of this is VO$_2$.  Below $T_*=340K$,
VO$_2$ becomes insulating with a gap of $\Delta=0.6$ eV. This ratio of
the gap to the critical temperature $\Delta/T_*$ is approximately
$20$.  This behavior should be contrasted with systems such as superconductors in which $U(1)$ symmetry is broken and $\Delta/T_c\approx1-2$.  That $\Delta/T_*$ well exceeds unity is one of the unresolved puzzles
with VO$_2$. It points to strong correlations being the source of the gap rather than the breaking of some spontaneous symmetry as in the case of superconductivity. Optical conductivity studies \cite{basov} reveal that spectral weight as far away as 6 eV contributes to the formation of the Drude peak at zero frequency once the Mott gap closes.  Such UV-IR mixing is a ubiquitous feature of Mott systems.  While we have argued that our holographic setup can capture the high-low energy spectral weight transfer, we have not yet addressed the finite temperature aspects of the Mott problem.  

\begin{figure}
\centering
\begin{tabular}{ccc}
\hskip -0.2in\includegraphics[width=45mm]{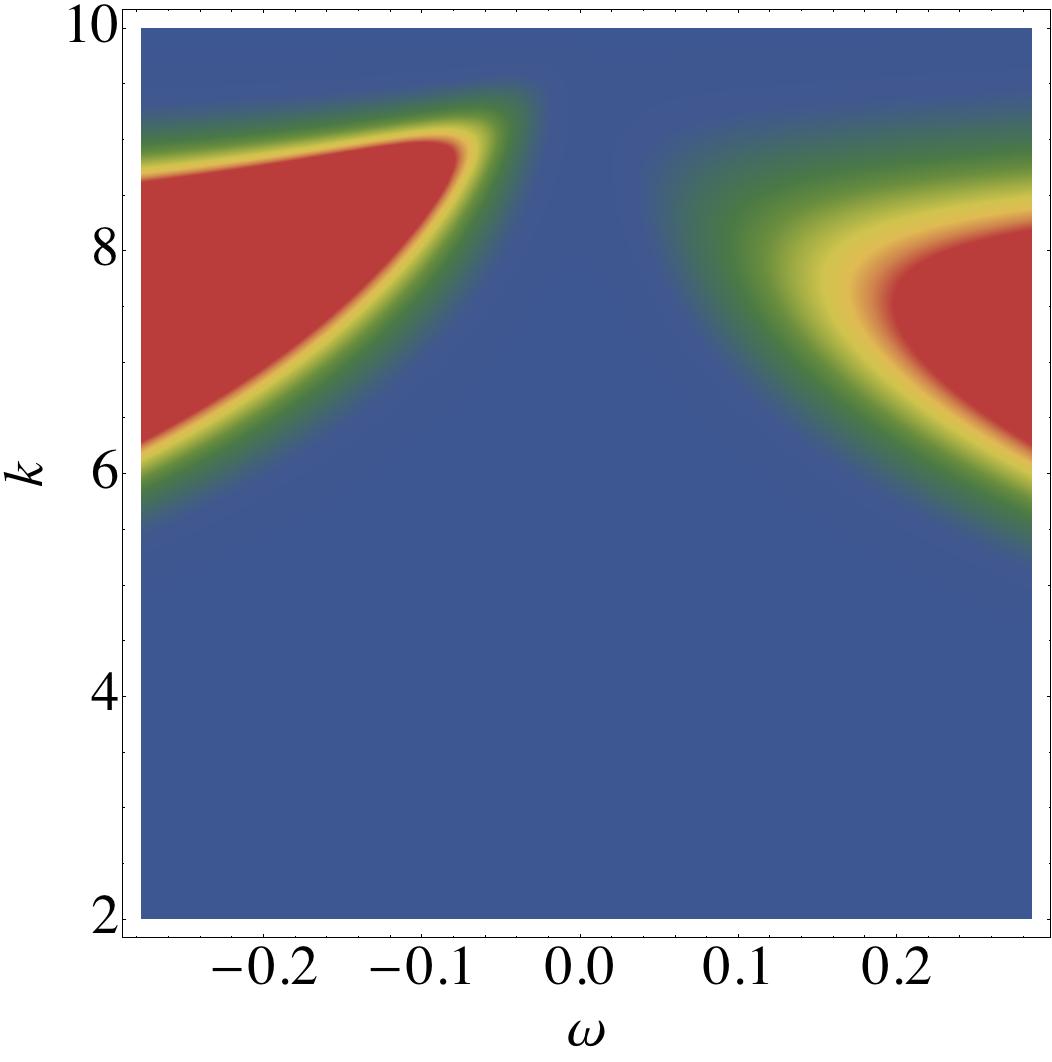} &
\includegraphics[width=45mm]{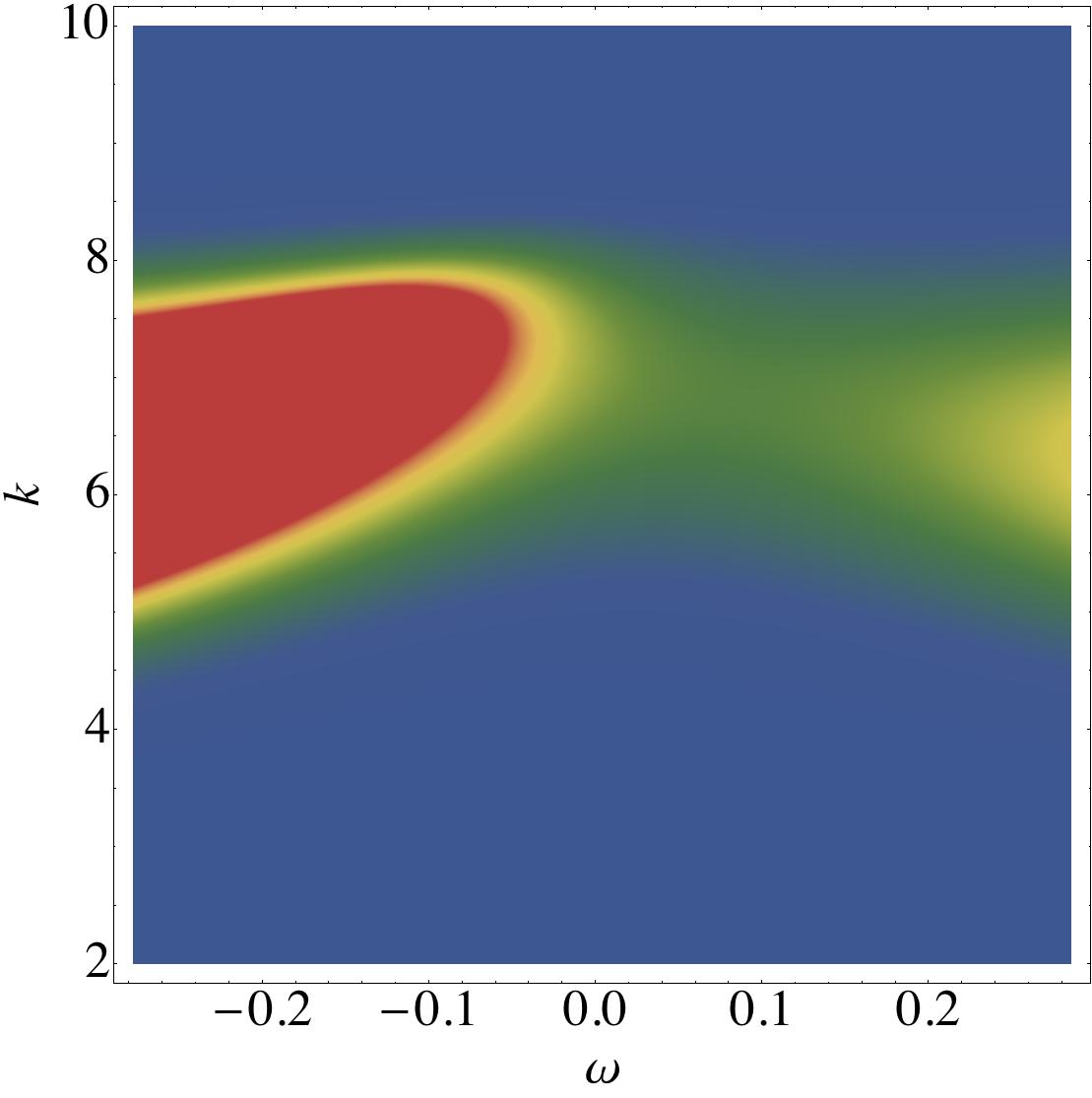} 
\end{tabular}
\caption{\footnotesize{A close-up of the density plots of ${\rm
      Im}\,G_{-}(\omega, k)$ for $ p=6$ and $T/\mu\simeq5.15\times
    10^{-3}$ (left) and $T/\mu\simeq 3.98\times 10^{-2}$  (right). A
    gap is still seen in the plot on the left while it is closed in the
    plot on the right. } }
\label{finiteT1}
\end{figure}

\begin{figure}
\centering
\includegraphics[width=75mm]{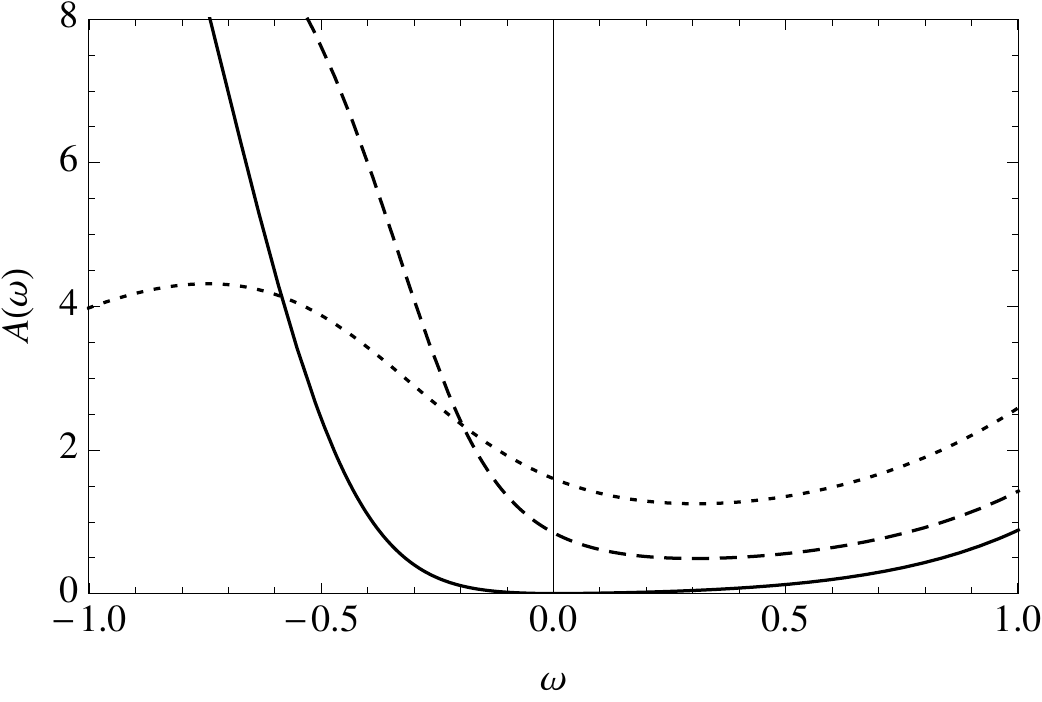} 
\caption{\footnotesize{A close-up of the density of states $A(\omega)$ at $p=6$ for  $T/\mu\simeq 0.44$ (dotted), $0.16$ (dashed) and $5.15\times 10^{-3}$ (solid).
%The gap at $T=0$ is gone at high enough temperatures. 
 } }
\label{finiteT2}
\end{figure}

The boundary theory we are investigating here can easily be
studied at finite temperature by considering the RN-AdS$_4$ background
away from extremality, namely for $0<Q< \sqrt{3}$.  Using the same
procedures outlined above, we obtained the spectral function and studied
the density of states as a function of temperature.  As
Figure \ref{finiteT1} reveals, the Mott gap observed here does in fact
close as the temperature increases.  Further, the transition is sharp.
To estimate the ratio of the zero-temperature gap to the temperature
at which the gap closes, $T_*$, we take a close-up of the density of
states and study its evolution as a function of 
temperature, see Figure \ref{finiteT2}.  Indeed for $p=6$ (or $p=7$), we find that
$\Delta/T_*\sim 10$.  Though smaller than $\Delta/T_*$ in vanadium
oxide, it does illustrate that dynamically generated gap we have found
here does possess non-trivial temperature dynamics. 

\begin{figure}
\centering
\hskip -0.05in\includegraphics[width=75mm]{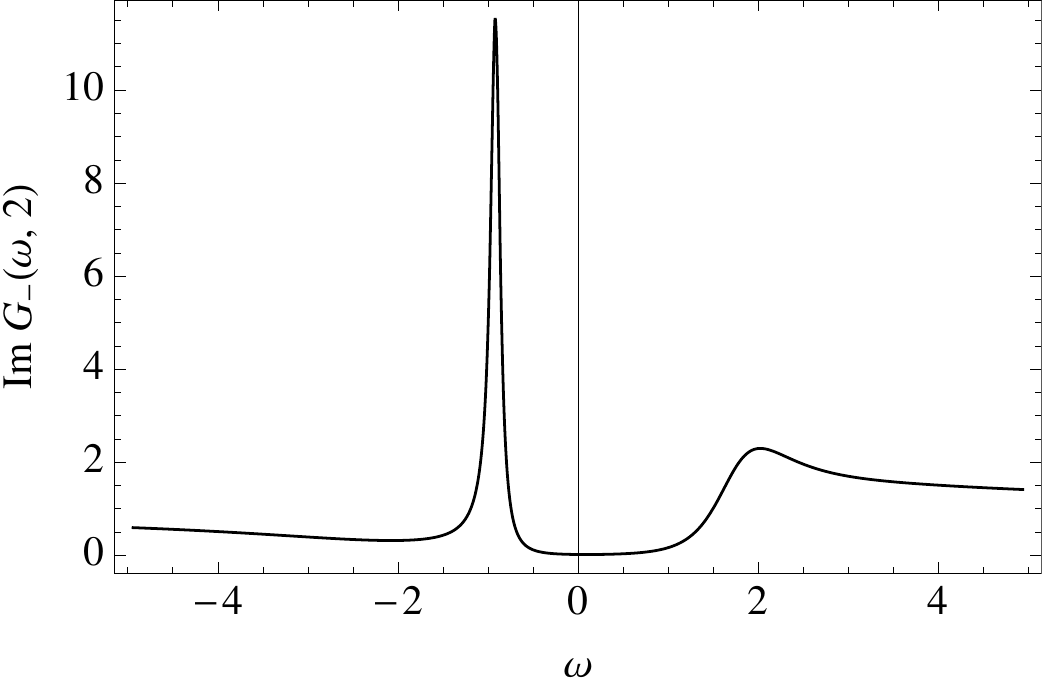} \\
\hskip -0.1in \includegraphics[width=75mm]{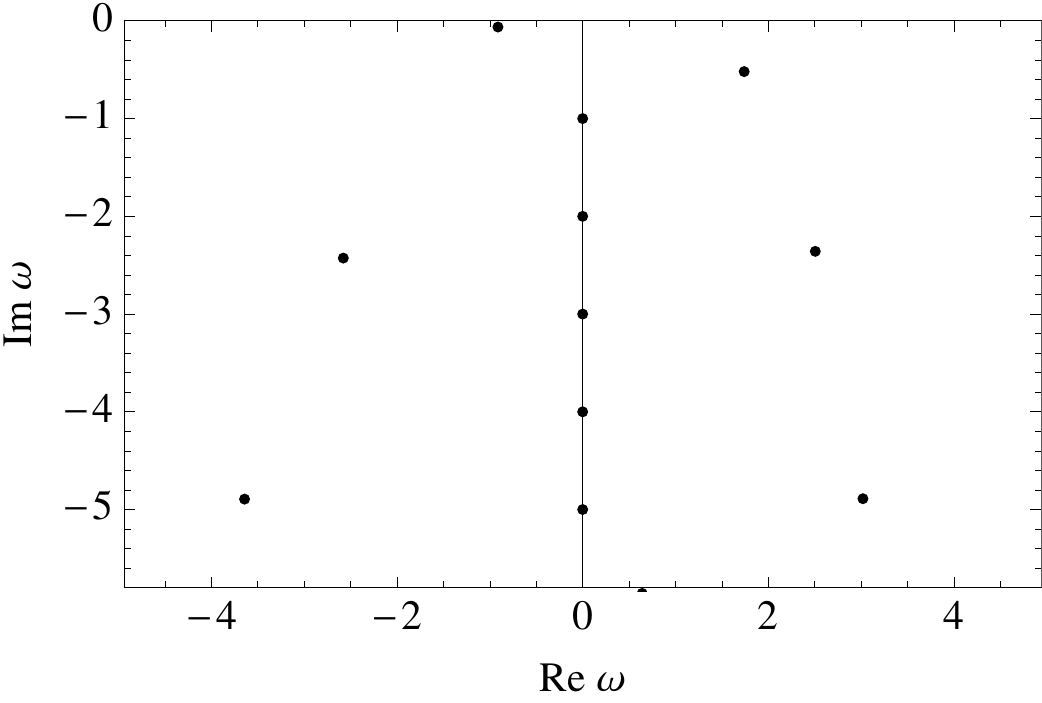} 
\caption{\label{ImGQNMFiniteT}\footnotesize{The plots in (a) and (b) show, for $k=2$ and $T/\mu = 0.16$, ${\rm Im}\,G_-(\omega, k)$ as a function of $\omega$ and the quasi-normal frequencies of $\alpha_-$, respectively.  Here, $d=3$, $p=5$, $q=1$, $m=0$, and $M=250$.} }
\end{figure}

For the record, we show in Figure \ref{ImGQNMFiniteT}  the quasi-normal frequencies of $\alpha_-$ (bottom plot) as well as ${\rm Im}\,G_{-}(\omega, k)$ (top plot) for $k=2$, $p=5$ and $T/\mu = 0.16$. Notice that the branch cut at zero temperature has dissolved  at finite temperature into a series of isolated poles on the negative imaginary axis\footnote{Note that in order to generate the bottom plot in Figure \ref{ImGQNMFiniteT}, the expression in \eqref{LeadingBehavioralpham} should be appropriately modified to reflect the fact that the system is at finite temperature. This is tied to the fact that at finite temperature $f(r)$ has a single zero at the horizon.}. Also, as it can easily be seen from the plots in Figure \ref{ImGQNMFiniteT}, the (real part of the) two quasi-normal frequencies of $\alpha_-$ which are closest to the real axis match quite well with the location of the peak on the left hand side and the bump on the right hand side in ${\rm Im}\,G_{-}(\omega, k=2)$.  The other quasi-normal frequencies  represent poles in ${\rm Im}\,G_{-}(\omega, k=2)$ which are relatively wide and have small residue. Although not shown, to the extent that we have checked, all the poles stay on the lower half $\omega$-plane as $k$ is varied.  

\section{Discussion}\label{sectionseven}

\begin{figure}
\includegraphics[width=70mm]{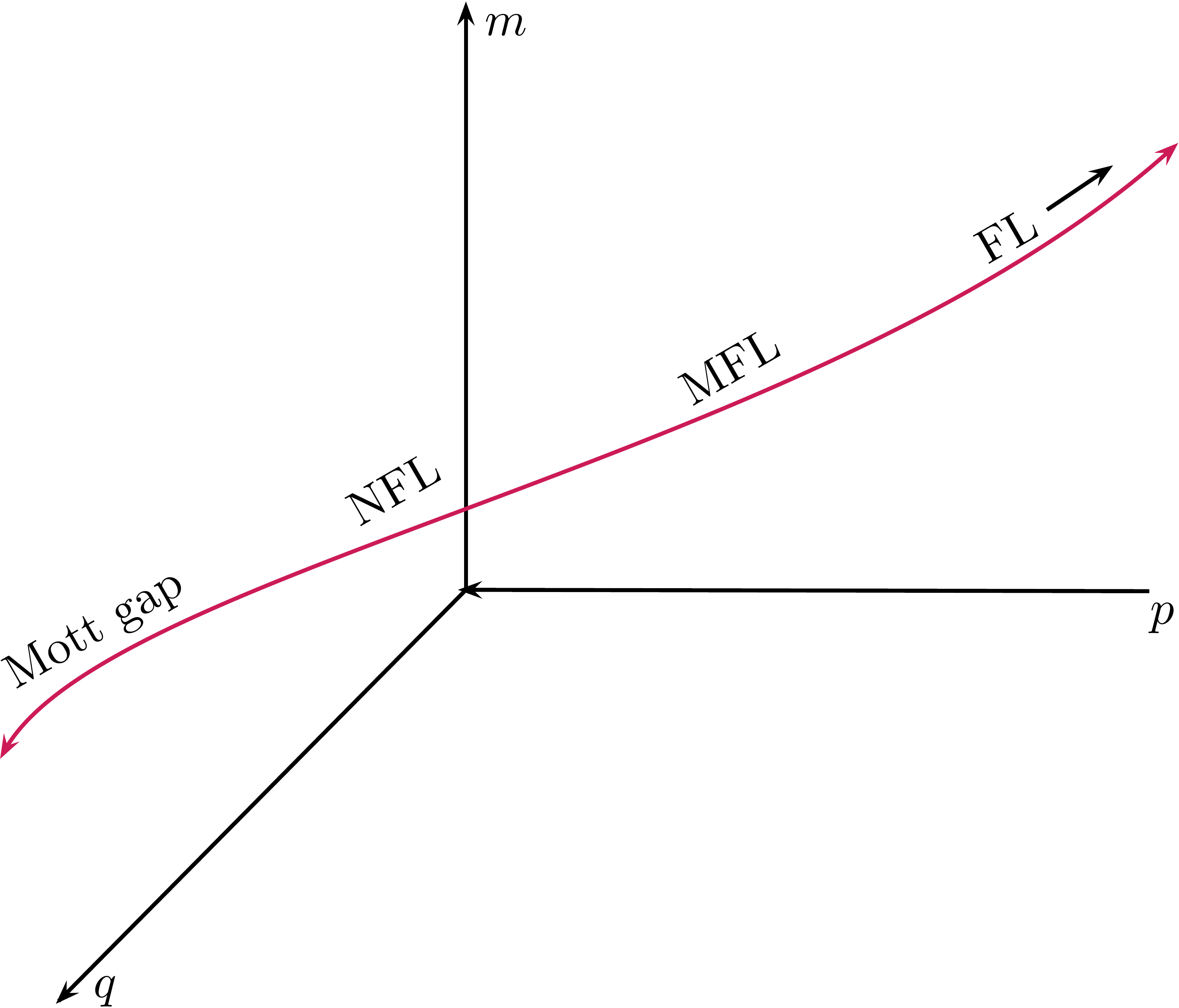} 
\caption{\label{cartoonPhase}\footnotesize{A cartoon of the zero temperature ``phase diagram" in the $m, q, p$ parameter space. Different regions of the phase diagram correspond to each of the principal structures in the cuprate phase diagram (compare to Figure \ref{pdiag}).
} }
\end{figure}
We have studied extensively the dichotomous behavior of the boundary theory 
fermion correlators in the presence of a bulk Pauli coupling in our
holographic set up.  For the boundary theory dual to the extremal RN-AdS$_4$ background, we showed that as we vary $p$ from large negative values up to a small positive value of $p=1/\sqrt{6}$ (while keeping  $m=0$ and $q=1$ fixed), the behavior of the excitations  change from Fermi liquid like (for $p<-0.53$), though not in the precise Landau sense in which the width of the excitations is quadratic in frequency, to a marginal Fermi liquid at $p=-0.53$ and on to a non-Fermi liquid for  $-0.53<p\leq1/\sqrt{6}$.  In the context of the earlier work \cite{Faulkner:2009wj} in which such behavior was observed by changing  the scaling dimension, as well as the charge, of the boundary theory fermion operator, the Pauli coupling offers a more direct connection with Mott physics.  Our argument here is based on the fact that for large positive values of $p$ a Mott gap arises (as evidenced by a vanishing of the quasiparticle residue shown in Figure \ref{ResGM}) in the spectrum of the boundary theory fermion operator without the apparent breaking of a continuous
symmetry. This is Mott physics. We have seen these features by holding
$m$ and $q$ fixed while varying $p$. It is clear that the basic
properties that we have seen will persist throughout a domain in the
$m,q,p$ parameter space. In Figure \ref{cartoonPhase}, we suggest that
some locus through the parameter space can be identified with doping
in the cuprate phase diagram: each of the principle features in the
normal state of the cuprates is present. It is interesting to compare
this heuristic phase diagram with that of the cuprates in which there is a continuous evolution from a
Mott insulator in the undoped state to a Fermi liquid in the overdoped
regime.  In between these extremes lie non-superconducting non-Fermi liquid states
characterized by a pseudogap (a suppression of the density of states
without any long-range superconductivity) and a strange metal in which
the resistivity is a linear function of temperature.  Our work suggests
that the Pauli coupling mimics the role of the electron filling.  

It is natural to investigate how the introduction of a superconducting condensate would complement the physics that we have discussed here. A suitable charged background at  zero-temperature was studied in \cite{Horowitz:2009ij}, following \cite{Hartnoll:2008vx, Hartnoll:2008kx}. For a range of parameters (namely, for $m_\phi^2-2q_\phi^2<-3/2$ where $m_\phi$ and $q_\phi$ are the mass and the charge of the bulk scalar field, respectively), a bulk solution with a non-zero charged scalar condensate is  preferred over the Reissner-Nordstr\"om solution. (Indeed, for this range of parameters, the Reissner-Nordstr\"om solution is unstable against turning on the scalar field in the bulk.) This solution is asymptotically AdS$_4$, and $m_\phi=0$ has a near horizon geometry that is also AdS$_4$ with a finite speed of light, $c_{{\rm IR}}$, and a finite dynamical exponent. The horizon of the zero-temperature solution (which is the Poincar\'e horizon of AdS$_4$) is at $r=0$ and there is no residual entropy at zero temperature.  Although we will discuss this elsewhere \cite{ellp}, preliminary studies of the effect of the Pauli coupling on the boundary theory fermion correlators\footnote{See \cite{Chen:2009pt,Faulkner:2009am, Gubser:2010dm, Ammon:2010pg,Benini:2010qc,Vegh:2010fc} where the authors analyze fermion correlators (in the absence of the Pauli interaction) in some superconducting backgrounds.} indicate that the main contributor to the gap in the fermion spectral density in the superconducting phase is the so-called Majorana scalar-fermion coupling (denoted by $\eta_5$ in \cite{Faulkner:2009am}). 
%(e.g. sufficiently large charge)
As there are limited tools available for the study
of ``electrons" at strong coupling, the model proposed here could offer
key insight into how superconductivity emerges from a background in
which all energy scales are coupled. 
%\begin{figure}
%\centering
%\includegraphics[width=6.5cm]{CupratesPhase.pdf}
%\caption{ \footnotesize{Heuristic phase diagram as a
%  function of holes doped into the copper-oxide plane.  The pseudogap
%  and strange metal are characterized by a depletion of the density of
%states and a $T$-linear resistivity, respectively. The pseudogap terminates at a zero-temperature critical point or quantum critical point (QCP). To the right is a
%Fermi liquid where weak-coupling physics is valid. }}
%\label{cupratepd}
%\end{figure}
\begin{figure}
%\vskip 0.5in
\centering
\begin{tabular}{ccc}
\hskip -0.15in\includegraphics[width=45mm]{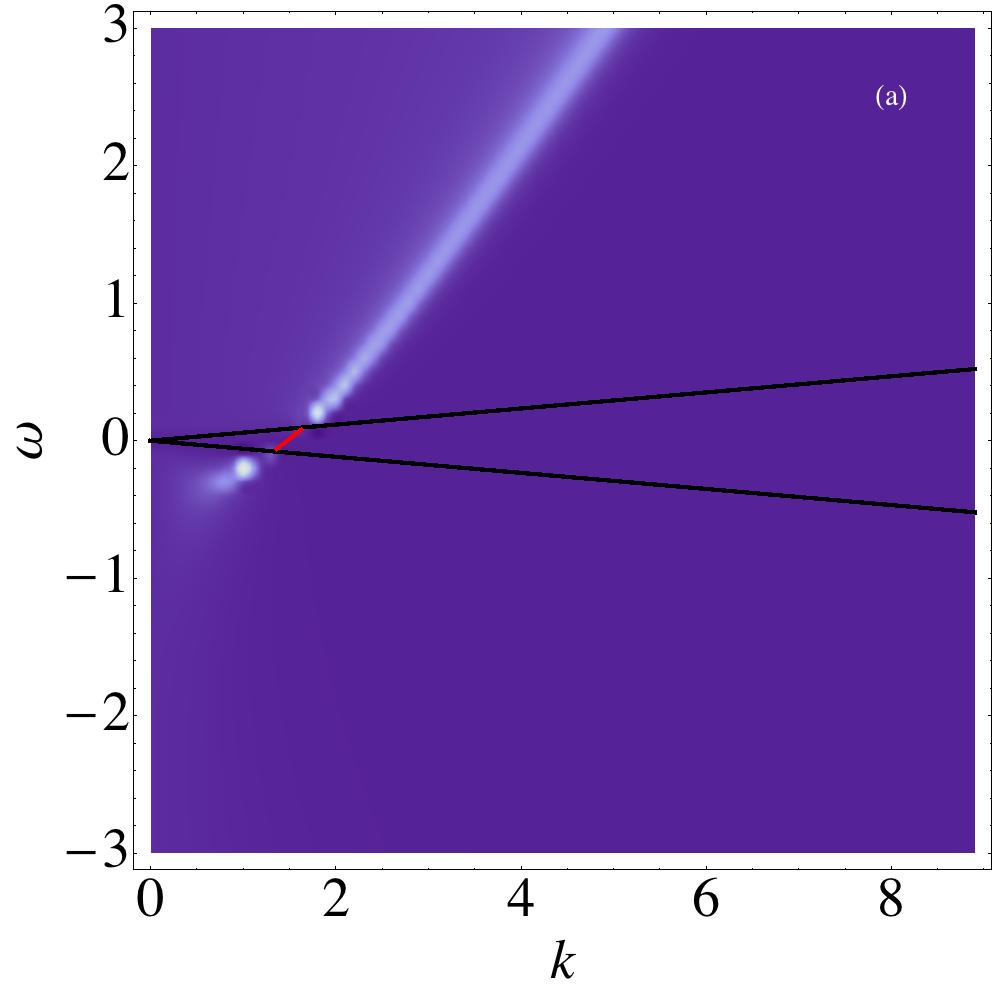} &
\includegraphics[width=45mm]{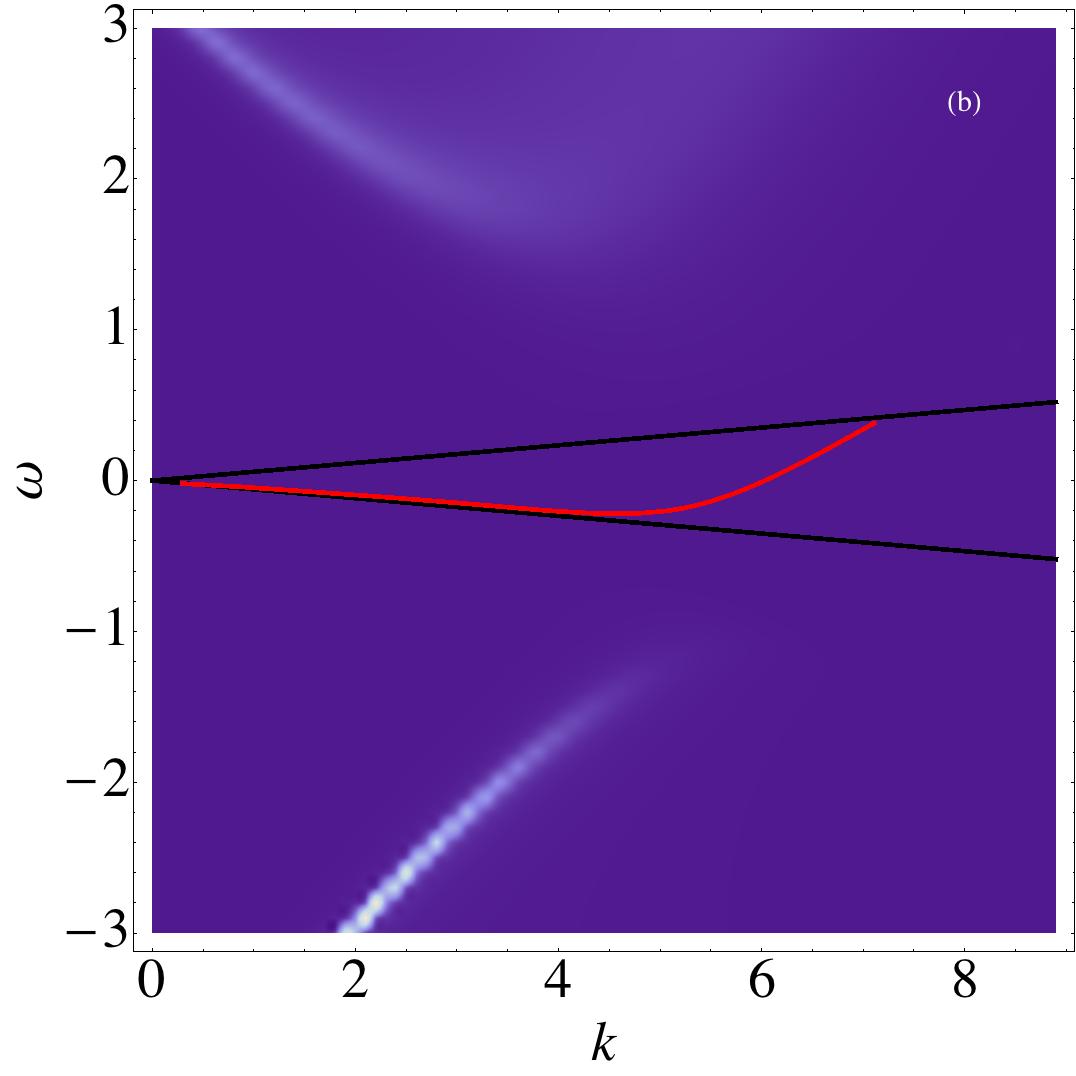} 
\end{tabular}
\caption{\label{SCDensityPlot} \footnotesize
{Density plot of the boundary theory fermion spectral function for (a) $p=0$ and (b) $p=3$. Here, $q_\phi=1.5$, $L=1$, and $\mu=2\sqrt{3}$. The black lines depict the  IR light-like region and the red curves represent the bound states.}}
\end{figure}
As a first step, we set $\eta_5=0$, and briefly discuss here the
effect of the Pauli coupling on fermion correlators in a boundary
theory dual to the superconducting background of
\cite{Horowitz:2009ij}. Suppose $m_\phi=0$, and $q_\phi>\sqrt{3}/2$,
so that the near horizon geometry is AdS$_4$ (with a characteristic
radius $L_{\rm IR}$) and assume there exists a (2+1)-dimensional IR
CFT dual to this AdS$_4$ near-horizon geometry. For definiteness, we
set $q_\phi=2q$, where $q$ is the charge of the bulk
fermion\footnote{This condition is not required when $\eta_5=0$. We
  consider this condition so that our analysis here can be generalized
  to the case where $\eta_5\neq0$ \cite{ellp}. Also, note that the
  convention of charge in \cite{Horowitz:2009ij, Faulkner:2009am} is
  different than our convention in previous sections by a factor of
  two, namely $q_{\rm here}=2q_{\rm there}$. In this discussion, we
  use the convention of \cite{Horowitz:2009ij, Faulkner:2009am} for
  $q$ and $q_\phi$.}.  The Dirac equation for $\psi_\pm$ (as well as
the corresponding flow equations for $\xi_\pm$) and the IR boundary
conditions  can easily be worked out. A crucial difference compared to
the case of the RN-AdS$_4$ is that here the dimension of the IR CFT
operators dual to $\psi_\pm (r\to 0)$ does not depend on
$p$. Depending on the sign of $s^2\equiv-\omega^2/c_{\rm IR}^2+k^2$,
the Green functions $G_\pm (\omega,k)$ exhibit different
behaviors. (Note that one still has $G_{-}
(\omega,k)=G_+(\omega,-k)$.) For $s^2>0$ (IR space-like region), the
horizon boundary conditions for $\xi_\pm$ are real and since the flow
equations are also real, one concludes that the boundary theory
fermion spectral density is zero in this region, except when there are
bound states (by which we mean poles of ${\rm Re}\, G_\pm (\omega,k)$
in the $s^2>0$ region) of the Dirac equation.  For $s^2<0$ (IR
time-like region), on the other hand, the horizon boundary conditions
are complex resulting generically in a non-vanishing fermion spectral
density. The IR space-like region is the primary feature of the finite
$c_{\rm IR}$ theory that distinguishes it from the RN-AdS$_4$
theory. In particular, there is no analogue here of the
log-oscillatory region (at $\omega=0$), and thus no analogous
mechanism for the disappearance of zero-frequency poles as $p$ is
increased. Indeed as we describe below, as $p$ is increased, the
zero-frequency pole moves but persists. Thus in
the presence of superconductivity where the U(1) symmetry is
spontaneously broken, the gap in the fermion spectral density is determined by the Majorana scalar-fermion coupling $\eta_5$. This implies that the parameter responsible for Mottness is distinct from those involved in superconductivity. 

\begin{figure}
\centering
\vskip 0.15in\includegraphics[width=7.5cm]{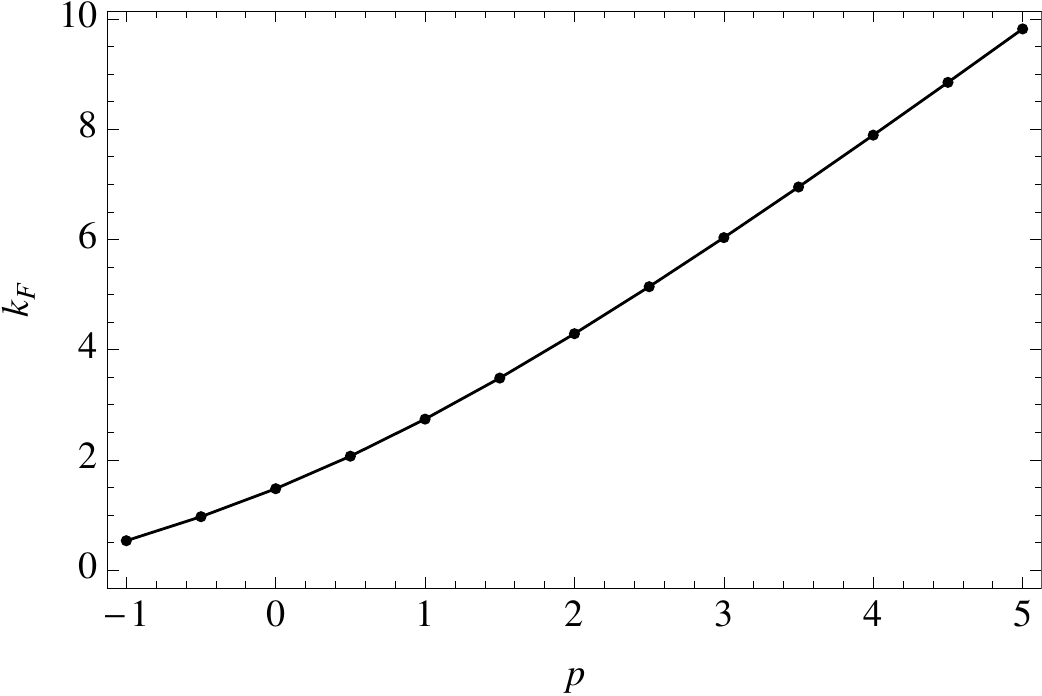} 
\caption{\label{SCFermiPeak}\footnotesize{$k_F$ as a function of $p$. Here, $q_\phi=1.5$, $L=1$, and $\mu=2\sqrt{3}$.} }
\end{figure}

Figure \ref{SCDensityPlot} shows a density plot of the fermion
spectral function for $p=0$ (left) and $p=3$ (right), where
$q_\phi=1.5$ and $\mu=2\sqrt{3}$.  The density plot for $p=0$, which
has been previously obtained in \cite{Faulkner:2009am}, is also shown
for the purpose of comparison with the density plot for a non-zero
value of $p$ such as $p=3$. As the plot in Figure
\ref{SCDensityPlot}(b) shows, turning on a non-zero value of $p$
suppresses the spectral density of the incoherent excitations (those
in the IR time-like region)  and pushes them away from the boundary
(the IR light-like region, depicted by solid black lines).  But, since
at $\omega=0$ there is a Fermi peak (and indeed an infinite number of
long-lived bound states for other values of $\omega$ in the IR
space-like region), turning on $p$ does not result in the formation of
a gap. In fact, for the above-mentioned parameters, we find that, for
$p=3$, $k_{\rm F}\approx 6.0$.  For $p=0$, the Fermi peak, for the
same parameters, is at $k_{\rm F}\approx 1.5$
\cite{Faulkner:2009am}. Turning on $p$ moves around the location of
this Fermi peak as shown in Figure \ref{SCFermiPeak}.  
%Consequently, the question is the fate of such peaks in the presence of the
%Majorana coupling.

We note again that there are regions of parameter space (in particular, $q_\phi$ and $m_\phi$) where the preferred geometry is either Reissner-Nordstr\"om or the superconducting geometry. Our results indicate that while the fermion gap in the superconducting geometry is controlled by the Majorana scalar-fermion coupling \cite{Horowitz:2009ij}, a Mott gap can still form in the Reissner-Nordstr\"om regime. The decoupling of these two effects is promising in the context of the cuprate phase diagram.

\section*{Acknowledgments}
We would like to thank E. Fradkin, and S. Hartnoll for discussions.  M.E. and P.W.P.  acknowledge financial support from the NSF DMR-0940992 and the Center for Emergent Superconductivity, a DOE Energy Frontier Research Center, Award Number DE- AC0298CH1088. R.G.L.\ is  supported by DOE grant FG02-91-ER40709 and would like to thank the Galileo Galilei Institute for Theoretical Physics for support and the participants of the program `AdS/CFT and the Holographic States of Matter' for many useful discussions.

%%%%%%%%%%%%%%%%%%%%%%%%%%%%%%%%%%%%%%%%%%%%%%%%
%\bibliographystyle{uiuchept}
%\bibliography{Fermibib}

\end{document}